\documentclass[11pt,american]{article}
\usepackage[T1]{fontenc}
\usepackage[latin9]{inputenc}
\usepackage{geometry}
\usepackage{color}
\usepackage{float}
\usepackage{amsmath}
\usepackage{amssymb}
\usepackage{graphicx}
\usepackage{esint}

\makeatletter
\usepackage{babel}

\usepackage{babel}

\usepackage{babel}

\usepackage{babel}

\usepackage{babel}

\usepackage{babel}

\usepackage{babel}

\@ifundefined{showcaptionsetup}{}{%
 \PassOptionsToPackage{caption=false}{subfig}}
\usepackage{subfig}
\makeatother

\usepackage{babel}
\begin{document}

\title{Non-equilibrium time evolution and rephasing in the quantum sine-Gordon
model}

\author{D. X. Horváth$^{1,3}$, I. Lovas$^{2,3}$, M. Kormos$^{1,3}$, G.
Takács$^{1,3}$ and G. Zaránd$^{2,3}$\\
 $^{1}${\small{}{}BME \textquotedbl{}Momentum\textquotedbl{} Statistical
Field Theory Research Group}\\
 {\small{}{}1111 Budapest, Budafoki út 8, Hungary}\\
 $^{2}${\small{}{}BME \textquotedbl{}Momentum\textquotedbl{} Exotic
Quantum Phases Research Group}\\
 {\small{}{}1111 Budapest, Budafoki út 8, Hungary}\\
 $^{3}${\small{}{}Department of Theoretical Physics, }\\
 {\small{}{}Budapest University of Technology and Economics}\\
 {\small{}{}1111 Budapest, Budafoki út 8, Hungary} }

\date{17th September 2018}

\maketitle
\global\long\def\ud{\mathrm{d}}
 \global\long\def\w{\omega}
 \global\long\def\fii{\varphi}
 \global\long\def\p{\partial}
 \global\long\def\lam{\lambda}
 \global\long\def\psid{\psi^{\dagger}}
\begin{abstract}
We discuss the non-equilibrium time evolution of the phase field in
the sine-Gordon model using two very different approaches: the truncated
Wigner approximation and the truncated conformal space approach. We
demonstrate that the two approaches agree for a period covering the
first few oscillations, thereby giving a solid theoretical prediction
in the framework of sine-Gordon model, which is thought to describe
the dynamics of two bosonic condensates in quasi-one-dimensional traps
coupled via a Josephson tunneling term. We conclude, however, that
the recently observed phase-locking behavior cannot be explained in
terms of homogeneous sine-Gordon dynamics, which hints at the role
of other degrees of freedom or inhomogeneity in the experimental system. 
\end{abstract}

\section{Introduction\label{sec:Intro}}

In recent years, understanding the out-of-equilibrium phenomena of
isolated quantum many-body systems has become a major challenge. The
recent experimental realization of such systems spurred considerable
interest and progress in the experimental and theoretical study of
non-equilibrium behavior. In particular, the use of cold atomic gases
led to controlled realizations of isolated quantum systems, and allowed
the observation of a number of astonishing non-equilibrium phenomena
\cite{NewtonCradle,ExperimentalNoThermalization1,ExperimentalNoThermalization3,GGEExperimental,ColdAtomSchm1,ColdAtomSchm2,Nagerl,Fukuhara,Kaufman}.
These include the lack of thermalization in quantum integrable systems
\cite{NewtonCradle,ExperimentalNoThermalization1,ExperimentalNoThermalization3,ExperimentalNoThermalization2}
or the experimental confirmation \cite{GGEExperimental} of the generalized
Gibbs ensemble (GGE) \cite{GGEProposal} as the valid description
of non-equilibrium steady states.

A particularly interesting experimental setup is provided by a bosonic
Josephson junction, consisting of two coupled superfluids in parallel
elongated traps \cite{Levy,Gati}. When the dynamics of the condensates
is dominated by continuum 1D physics, the relative phase of the condensates
can be described by the sine-Gordon model \cite{gritsev}. In thermal
equilibrium and under suitable conditions, this fact was demonstrated
experimentally in Ref.~\cite{Schmiedmayer} by comparing the measured
correlations to the prediction of classical thermal sine-Gordon model
\cite{Beck}. On the other hand, the out-of-equilibrium behavior of
the system of the coupled condensates was found to display intriguing
behavior such as a rapid phase-locking \cite{SchmiedmayerPhase},
so far unexplained from the dynamics of the sine-Gordon field theory.

In this work, we analyze this rephasing phenomenon within the theoretical
framework of the homogeneous sine-Gordon model through a combination
of two powerful though approximate methods: the Truncated Wigner Approximation
(TWA) and the Truncated Conformal Space Approach (TCSA). We compare
these two approaches in the strongly interacting regime, where we
find an excellent agreement between them. However, our TWA results
in the weakly interacting regime, relevant for the experiments, clearly
disagree with experimental observations, thereby leading us to the
conclusion that the homogeneous sine-Gordon model is \emph{insufficient}
to account for the experimental observations.

The sine-Gordon model has attracted interest since long \cite{Thirring,Coleman,Mandelstam},
and is considered to be a paradigmatic example of an integrable quantum
field theory \cite{Faddeev,Sklyanin,ZamZam}. Due to integrability,
many quantities can be computed exactly, such as the scattering amplitudes
\cite{ZamZam}, exact expectation values \cite{LukZam}, and form
factors \cite{Smirnov}. Integrability allows the application of powerful
methods to compute the long-distance expansion of zero-temperature
two-point correlators \cite{Essler}, although only very limited results
are available on quantum correlation functions under more general
conditions, and they are mostly confined to one-point functions in
thermal equilibrium \cite{Buccheri,Hegedus}.

Therefore numerical and approximate methods are of great value and
must also be invoked to understand these non-equilibrium systems.
One possibility is to resort to a semiclassical description, which
allows the construction of one- and two-point functions using a quasi-particle
picture \cite{DamleSachdev,Kormos}. Semiclassical methods can, however,
only partially account for the quantum dynamics.

Here we apply an alternative approach, the so-called truncated Wigner
approximation (TWA) \cite{Torre,polkovnikov,polkovnikov2}, giving
a slightly different quasi-classical approximation of the phase dynamics.
Unfortunately, it is very hard to control the accuracy of TWA. TWA
has so far only been validated by perturbation theory (that captures
only the first peak in the evolution) and by a scaling law that simply
follows from conformal field theory considerations. TWA predicts an
interesting oscillation with slowly decaying amplitude at longer times
\cite{Torre}. However, as shown here, quantum corrections become
dominant in this long time regime, and the TWA approximation becomes
uncontrolled.

To validate the truncated Wigner approach at these longer times, we
resort to the truncated conformal space approach (TCSA). This approach
was originally introduced by Yurov and Zamolodchikov to describe the
finite volume spectrum of two-dimensional QFTs \cite{TCSAYurovZamo}
, and it can be also used to describe non-equilibrium time evolution
in quantum field theories \cite{QuenchTCSA}, initial state overlaps
\cite{sGOverlaps} and multi-point correlation functions in and out
of equilibrium \cite{Kukuljan}.

Here we use the TCSA and the TWA approaches to examine phase locking
during the first few oscillations with two initial conditions. The
first corresponds to preparing two identical condensates independently
in their ground state, and switching on tunneling at time zero. The
other initial condition differs by preparing the two condensates with
a well-defined difference of the atom numbers in the respective trap.
For technical reasons, here we focus on homogeneous 1D systems with
periodic boundary condition and neglect density inhomogeneities. Both
initial conditions yield weakly damped oscillations even in the strongly
interacting regime. The excellent agreement between TCSA and TWA provides
a strong validation of the results presented here.

Before considering the sine-Gordon model in its full glory, we can
gain a rough understanding of the dynamics by considering a single
mode model of the condensates, corresponding to the quantum pendulum
\begin{equation}
H_{{\rm pendulum}}=Un_{0}^{2}-JN\cos\varphi_{0},\label{eq:pendulum}
\end{equation}
where the canonical conjugate variables $\varphi_{0}$ and $n_{0}$
satisfy $[\varphi_{0},n_{0}]=i$. Hamiltonian \eqref{eq:pendulum}
describes the time evolution of the relative phase $\varphi_{0}$
and particle number difference $n_{0}$. Here $U$ characterizes the
interaction between the atoms in the same condensate, $J$ denotes
the tunnel coupling between the potential wells, $N$ is the total
number of atoms, and Eq. \eqref{eq:pendulum} is valid in the regime
of small particle number difference $n_{0}\ll N$.

\begin{figure}[t!]
\centering{}\includegraphics[width=0.4\columnwidth]{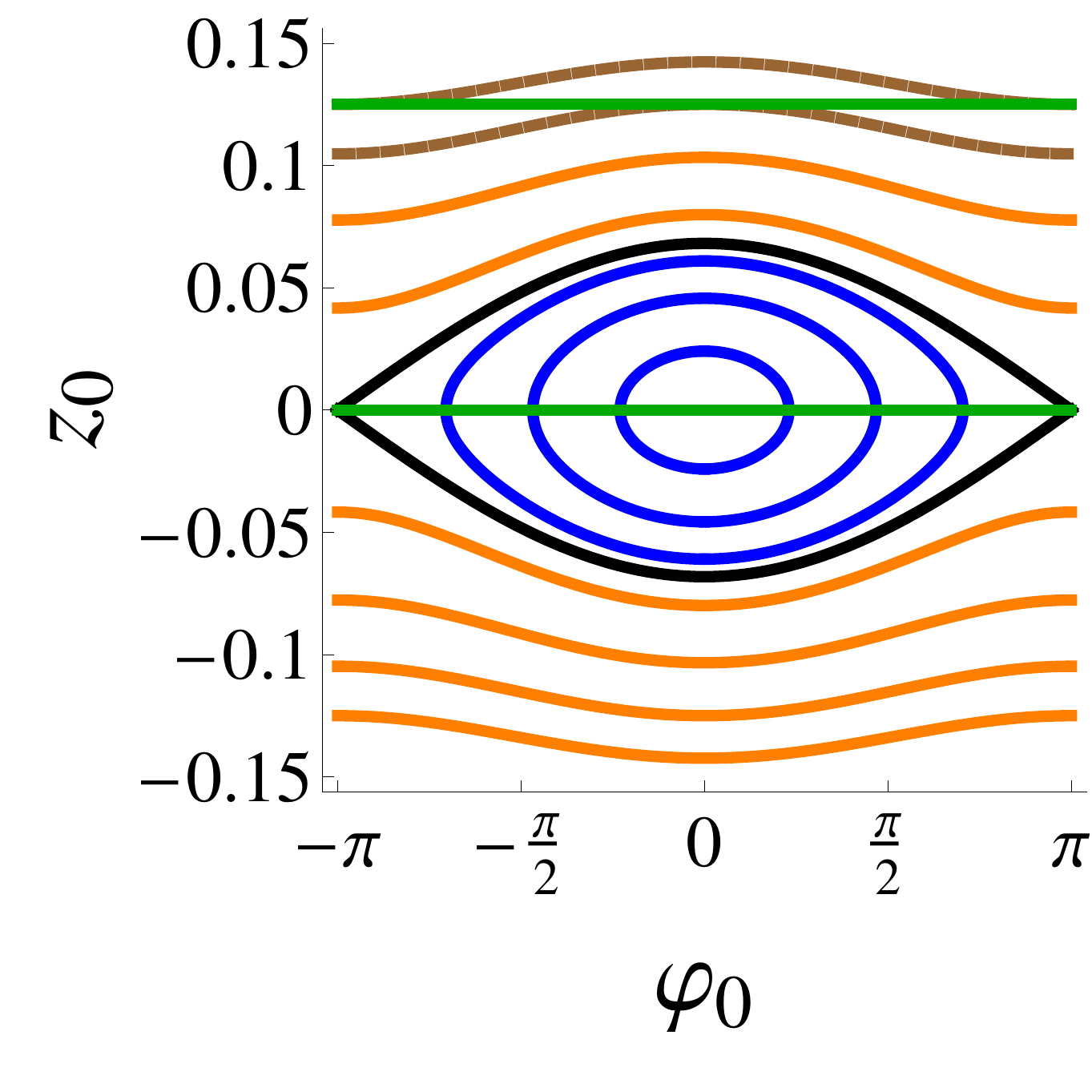} \caption{Classical trajectories of the pendulum model \eqref{eq:pendulum},
plotted in terms of phase $\varphi_{0}$ and rescaled particle number
difference $z_{0}=2\,n_{0}/N$. The separatrix (black) separates the
self-trapped regime with trajectories constrained to $z_{0}>0$ or
$z_{0}<0$ half-plane (orange), from the trajectories visiting both
half-planes (blue). The initial states considered in Sec. \ref{sec:TimeEvol},
corresponding to a well-defined particle number difference and a random
phase, can be visualised as horizontal lines (green, dashed). In the
self-trapped regime, the small frequency shift between the trajectories
cut by this line (brown) amounts to beating phenomena.}
\label{fig:pendulum} 
\end{figure}

Although for the quasi one dimensional condensates considered in this
paper, fluctuations \textendash{} incorporated in the sine-Gordon
model \textendash{} play an essential role, and the single mode approximation
Eq. \eqref{eq:pendulum} breaks down, the dynamics of the pendulum
model \eqref{eq:pendulum}, nevertheless, still offers a qualitative
understanding of the time evolution under the sine-Gordon Hamiltonian.
We display the classical trajectories of the pendulum \eqref{eq:pendulum}
in Fig. \ref{fig:pendulum}, determined by energy conservation 
\begin{equation}
\dfrac{UN^{2}}{4}z_{0}^{2}-JN\cos\varphi_{0}=E,
\end{equation}
with $z_{0}=2\,n_{0}/N$. For large enough $E$, the trajectories
are confined to a self-trapped domain, constrained to the half-plane
$z_{0}>0$ or $z_{0}<0$, since the high interaction energy prevents
leveling off the number of particles in the two potential wells. This
region is separated from the low energy trajectories, visiting both
half planes, by a separatrix.

Both initial states considered in this work correspond to a well defined
particle number difference $n_{0}$, and to a uniformly distributed
random phase $\varphi_{0}$; they are visualized as dashed horizontal
lines in Fig. \ref{fig:pendulum}. The non-equilibrium dynamics can
be qualitatively understood in terms of the classical trajectories
intersected by these horizontal lines. To test TWA and TCSA in both
phases, first we consider two identical condensates in Sec. \ref{sec:TimeEvolI},
corresponding to $n_{0}=0$, where the dynamics is determined by the
classical trajectories lying in the non-trapped phase. Then in Sec.
\ref{sec:Time-evolution-II} we choose a large particle number imbalance
$n_{0}$, such that all relevant trajectories are self-trapped. Unfortunately,
as we discuss below, for technical reasons we cannot compare the TCSA
and TWA methods for initial states intersecting the separatrix. Nevertheless,
for initial states far enough from the boundary of self-trapping,
we find an excellent agreement between TWA and TCSA methods in both
phases. Moreover, in the self-trapped phase all intersected trajectories
lie in a narrow frequency window. As we discuss in Sec. \ref{sec:Time-evolution-II},
the non-equilibrium expectation values considered here oscillate with
the typical frequency of these classical trajectories, while the small
frequency shift between the relevant trajectories gives rise to beating
effects.

The outline of the paper is as follows. Sec.~\ref{sec:sGmodel} reviews
the necessary ingredients of the sine-Gordon description of the coupled
quasi-1D condensates. In Sec.~\ref{sec:HLattice-HQFT} we provide
the detailed connections between the lattice regularized description
used in the TWA and the perturbed conformal field theory framework
of the TCSA, which allows for their detailed numerical comparison.
Sec. \ref{sec:TWA_TCSA} outlines the description of both the TWA
and TCSA methods themselves, while the results for the two initial
conditions are presented in Secs. \ref{sec:TimeEvolI} and \ref{sec:Time-evolution-II},
respectively. Finally, our conclusions are presented in Sec. \ref{sec:Conclusions}.
Certain technicalities are relegated to appendices: App. \ref{app:sGmap}
provides more details on the mapping of the coupled condensates to
a sine-Gordon model, App. \ref{app:normal} contains the technical
details on normal ordering needed to compare observables between TWA
and TCSA, App. \ref{app:cutoff} describes the extrapolation procedure
used in TCSA to eliminate the leading truncation effects, while App.
\ref{app:twa} reviews the derivation of the TWA and its leading quantum
correction.

\section{Sine-Gordon description of ultracold one-dimensional bosons}

\label{sec:sGmodel}

The physics of two Josephson-coupled one-dimensional interacting quasi-condensates
can be described to a very good approximation by the sine-Gordon model.
For a precise mapping, one usually considers two quasi one-dimensional
gases, described by the Hamiltonians \cite{Cazalilla,CoupledCondensates}
\begin{equation}
H_{0}=\sum_{j=1,2}\int\ud x\left\{ \frac{\hbar^{2}}{2m}\p_{x}\psi_{j}^{\dag}(x)\p_{x}\psi_{j}(x)+\frac{g}{2}\psi_{j}^{\dag}(x)\psi_{j}^{\dag}(x)\psi_{j}(x)\psi_{j}(x)+[V(x)-\mu]\psi_{j}^{\dag}(x)\psi_{j}(x)\right\} ,\label{eq:LiebLiniger}
\end{equation}
coupled by the Josephson tunneling term, 
\begin{equation}
H_{J}=-J\int\ud x\left[\psi_{1}^{\dag}(x)\psi_{2}(x)+\psi_{2}^{\dag}(x)\psi_{1}(x)\right]\,.\label{HMicroscopic}
\end{equation}
Here $\psi_{1}(x),\psi_{2}(x)$ denote the bosonic fields of the two
quasi-condensates, $V(x)$ is the longitudinal trap potential, $J$
the tunneling amplitude, and $g$ stands for the effective one-dimensional
interaction \cite{ZwergerReview}. In the rest of this paper, we shall
neglect the trapping potential and focus on homogeneous condensates.
Furthermore, since boundary conditions do not influence the dynamics
discussed in an essential way, we shall impose periodic boundary conditions
for the sake of simplicity.

The value of the coupling $g$ depends sensitively on the shape of
the transverse trapping potential, and can be approximated as \cite{Olshanii}
\begin{equation}
g\approx\frac{2\hbar^{2}a_{s}}{ml_{\perp}^{2}}\left(1-1.036\frac{a_{s}}{l_{\perp}}\right)^{-1}\;,
\end{equation}
with $l_{\perp}=\sqrt{\hbar/(m\w_{\perp})}$ the transverse oscillator
length associated with the frequency $\w_{\perp}$ of the radial confining
potential, and $a_{s}$ the three-dimensional $s$-wave scattering
length of the atoms. For weak interactions, $a_{s}\ll l_{\perp}$,
one simply obtains $g\approx2\hbar\omega_{\perp}a_{s}.$

To describe this interacting system, one often refers to \textquotedbl{}bosonisation\textquotedbl{}~\cite{Haldane},
and represents the trapped bosons in terms of their phase $\varphi_{j}$
and density $\rho_{j}$ as 
\begin{equation}
\psi_{j}(x)=\sqrt{\rho_{j}(x)}\;e^{i\varphi_{j}(x)}\;,\label{eq:phase_rep}
\end{equation}
with the density fluctuations $\delta\rho_{j}(x)$ and the phase $\varphi_{j}(x)$
playing the role of conjugate variables. Substituting \eqref{eq:phase_rep}
into Eq.~\eqref{eq:LiebLiniger}, one obtains a hydrodynamical description
of the condensates. To leading order in the density and phase fluctuations,
the relative phase $\varphi_{r}=\varphi_{2}-\varphi_{1}$ decouples
from the total phase, $\varphi_{1}+\varphi_{2}$ and the dynamics
of the former is described by the sine-Gordon Hamiltonian, 
\begin{equation}
H_{\mathrm{r}}=\frac{\hbar c}{2}\int\ud x\left\{ \frac{\pi}{K}\Pi_{r}^{2}+\frac{K}{\pi}\left(\partial_{x}\varphi_{r}\right)^{2}\right\} -2J\rho_{0}\int\ud x\cos\varphi_{r}\,,\label{eq:H_r}
\end{equation}
with $\Pi_{r}$ the relative canonical momentum, $c$ the sound velocity
and $K$ the so-called Luttinger parameter. For the homogeneous gas
considered here the average density $\rho_{0}$ is related to the
total number of atoms $N$ confined the length $L$ of the system
as $\rho_{0}=N/2L$. We note that $K$ is a rather non-trivial function
of the dimensionless interaction, $\gamma={mg}/{(\hbar^{2}\rho_{0})}$
(see Appendix~\ref{sec:sGMapping}); for weakly interacting bosons
$K\gg1$, while very strong atom-atom interactions yield $K\approx1/2$
.

The usual, field theoretical form of the sine-Gordon model can be
obtained by setting $\hbar=c=1$ and rescaling the relative field
and momentum as ${\color{black}\sqrt{\frac{K}{\pi}}\varphi_{r}\to\phi}$
and ${\color{black}\sqrt{\frac{\pi}{K}}\Pi_{r}\to\Pi}$, yielding
\begin{equation}
H_{\mathrm{sG}}=\frac{\hbar c}{2}\int\ud x\,\left\{ \Pi^{2}+\left(\partial_{x}\phi\right)^{2}\right\} -\frac{\mu^{2}}{\beta^{2}}\int\ud x\cos\left(\beta\phi\right)\,,\label{sGHamiltonianFirst}
\end{equation}
with the interaction parameter $\beta$ related to the Luttinger parameter
$K$ as\textbf{ 
\begin{equation}
\beta=\sqrt{\frac{\pi}{K}}\:,\label{eq:betaK}
\end{equation}
}and with $\mu=\sqrt{2\pi J\rho_{0}/K}.$

\section{Lattice regularized and perturbed conformal Hamiltonians for the
sine-Gordon model \label{sec:HLattice-HQFT}}

In order to make the sine-Gordon theory described by the Hamiltonian
\eqref{sGHamiltonianFirst} well-defined, it is necessary to introduce
some regularization scheme. In this section, we discuss two prescriptions
that are directly related to the TWA and TCSA methods. Sec. \ref{subsec:LatticeReg}
is devoted to reviewing the lattice regularization of the model, while
in Sec. \ref{subsec:PertCFT} the theory is formulated as the relevant
perturbation of a massless bosonic conformal field theory. The parameters
entering the two different formulations are matched later in Sec.
\ref{sec:TWA_TCSA}, after a description of the TWA and TCSA approaches.

\subsection{Lattice regularization\label{subsec:LatticeReg}}

The first regularization uses a spatial discretization by a lattice
with spacing $a$. To keep a direct connection with the cold atomic
system we use parameters and variables corresponding to Eq. \eqref{eq:H_r}.
Denoting $\varphi=\varphi_{r}$, the lattice regularization of $H_{\mathrm{r}}$
in \eqref{eq:H_r} can be written as 
\begin{equation}
H_{\mathrm{Lat}}=\frac{\hbar c}{2}\sum_{j=1}^{N_{s}}\left(\frac{\pi}{Ka}n_{j}^{2}+\frac{K}{\pi a}\left(\varphi_{j}-\varphi_{j-1}\right)^{2}\right)-2J\rho_{0}a\sum_{j=1}^{N_{s}}\cos\varphi_{j}\;,\label{eq:HLat}
\end{equation}
where 
\[
n_{k}=\frac{Ka}{\pi c}\dot{\varphi}_{k}\quad,\quad\left[\varphi_{j},n_{k}\right]=i\delta_{jk}.
\]
The number of sites is given by $N_{s}=L/a$ and periodic boundary
conditions are assumed. This scheme is natural in view of the original
microscopic Hamiltonian (\ref{HMicroscopic}), as the bosonisation
formula \eqref{eq:phase_rep} is written in terms of a coarse grained
density and phase, resulting in an effective long wave length description
valid above a short-distance cut-off, the so-called healing length
(see Appendix \ref{app:sGmap}). In the non-interacting case $J=0$
the spectrum of the lattice Hamiltonian \eqref{eq:HLat} can be expressed
as

\begin{equation}
\varepsilon_{k}=\frac{2\hbar c}{a}\left|\sin\frac{ka}{2}\right|\,,\label{eq:LatticeDispRel}
\end{equation}
reducing to the linear spectrum $\varepsilon_{k}=\hbar ck$ of a Luttinger
liquid for small wave numbers $k\ll1/a$.

Notice that the cosine term in \eqref{eq:HLat} is not normal ordered.
To treat it semiclassically, normal ordering with respect to the bosonic
vacuum is necessary. As shown in App. \ref{sec:CosFi}, this amounts
to an additional prefactor

\begin{equation}
\cos\varphi_{i}=\mathcal{N}:\cos\varphi_{i}:\,,
\end{equation}
with

\begin{equation}
\mathcal{N}=\exp\left(-\frac{\pi\Delta}{N_{s}}\right)\prod_{n=1}^{N_{s}/2-1}\exp\left(-\frac{2\pi\Delta}{N_{s}\sin\frac{\pi n}{N_{s}}}\right)\,,
\end{equation}
where

\begin{equation}
\Delta=\frac{1}{8K}=\frac{\beta^{2}}{8\pi}\:.\label{eq:delta}
\end{equation}
The lattice regularization of the Hamiltonian \eqref{eq:H_r} finally
takes the form

\begin{equation}
H_{\mathrm{Lat}}=\frac{\hbar c}{2}\sum_{i}\left(\frac{\pi}{Ka}n_{i}^{2}+\frac{K}{\pi a}\left(\varphi_{i}-\varphi_{i-1}\right)^{2}\right)-2J\rho_{0}a\,\mathcal{N}\sum_{i}:\cos\varphi_{i}:\,.\label{eq:Hlat}
\end{equation}

\subsection{Perturbed conformal field theory formulation\label{subsec:PertCFT}}

A paradigmatic approach to a massive quantum field theory is to regard
it as a perturbation of an ultra-violet (UV) conformal field theory
(CFT) \cite{BPZ} with appropriate relevant operators. In this terminology,
perturbation does not mean that the coupling is considered to be weak.
Rather, it is understood as a deformation of the conformal field theory.
Indeed, in models with one space dimension, there exist powerful non-perturbative
methods which allow for the treatment of these models at strong coupling
as well. In this subsection we use the usual convention of conformal
field theory, and work in units $\hbar=1$ and $c=1$.

For the sine-Gordon model the corresponding description treats \eqref{sGHamiltonianFirst}
as a compactified massless bosonic conformal field theory, perturbed
by the relevant operator $\int\ud x\,:\cos\beta\phi:$, where compactification
of the bosonic field $\phi$ means that it takes values on a circle
with the identification $\phi\equiv\phi+m\frac{2\pi}{\beta}$, and
space-time has a cylindrical geometry due to periodic boundary conditions
(PBC) $x\equiv x+L$. Then the perturbed conformal Hamiltonian $H_{PCFT}$
reads

\begin{align}
H_{\mathrm{PCFT}} & =\int_{0}^{L}\ud x\,\frac{1}{2}:\left(\partial_{t}\phi\right)^{2}+\left(\partial_{x}\phi\right)^{2}:-\frac{\lambda}{2}\int_{0}^{L}\ud x\,\left(V_{1}^{\mathrm{cyl}}+V_{-1}^{\mathrm{cyl}}\right)\:,\label{eq:pcft_Hamiltonian}
\end{align}
where the exponential fields

\begin{equation}
V_{n}^{\mathrm{cyl}}=:e^{in\beta\phi}:^{\mathrm{cyl}}\label{eq:vertexops}
\end{equation}
are called vertex operators, and the semicolon denotes normal ordering
with respect to the massless scalar field modes. The upper index ``cyl''
of the normal ordering indicates that these vertex operators have
a canonical CFT normalization specified below in \eqref{eq:vertexop_norm},
and acquire an anomalous dimension. As a result, the coupling $\lambda$
in the Hamiltonian \eqref{eq:pcft_Hamiltonian} has a nontrivial dimension
related to the scaling exponent $\Delta$ \eqref{eq:delta}. Integrability
allows to determine its exact relation to the mass gap \cite{sGMassGap}:

\begin{equation}
\lambda=\left(2\sin\frac{\xi\pi}{2}\right)^{^{2\Delta-2}}\frac{2\Gamma(\Delta)}{\pi\Gamma(1-\Delta)}\left(\frac{\sqrt{\pi}\Gamma\left(\frac{1}{2-2\Delta}\right)m_{1}}{2\Gamma\left(\frac{\Delta}{2-2\Delta}\right)}\right)^{2-2\Delta}\qquad\text{with}\qquad\xi=\frac{\beta^{2}}{8\pi-\beta^{2}}\:,\label{eq:massgap}
\end{equation}
where $m_{1}$ is the mass of the first breather in the spectrum sine-Gordon
QFT, which is the lightest neutral excitation in the attractive regime\footnote{In the repulsive regime $4\pi\leq\beta^{2}\leq8\pi$ the lightest
excitations are topologically charged solitons, whose mass is similarly
related to $\lambda;$ the point $\beta^{2}=8\pi$ corresponds to
a Kosterlitz-Thouless transition above which the cosine perturbation
becomes irrelevant and the spectrum is gapless. } i.e. for $\beta^{2}<4\pi$. Relation \eqref{eq:massgap} allows one
to express all physical quantities in units of appropriate powers
of the first breather mass $m_{1}$.

Following the usual CFT procedure, the theory is continued analytically
to imaginary time $\tau=-it$, and we introduce complex coordinates
$w=\tau-ix$, $\bar{w}=\tau+ix$ on the resulting Euclidean space-time
cylinder. The normalization of the vertex operators is then specified
by the following short distance behavior of their two-point functions:
\begin{equation}
\langle0|V_{n}^{\mathrm{cyl}}(w_{1},\bar{w}_{1})V_{m}^{\mathrm{cyl}}(w_{2},\bar{w}_{2})|0\rangle=\frac{\delta_{n,-m}}{|w_{1}-w_{2}|^{4n^{2}\Delta}}+\text{subleading terms}\,.\label{eq:vertexop_norm}
\end{equation}
This shows explicitly that the vertex operators $V_{n}^{\mathrm{cyl}}$
have dimensions of $\left(\text{length}\right)^{-2n^{2}\Delta}$,
which in units $\hbar=1=c$ is the same as $\left(\text{energy}\right)^{2n^{2}\Delta}$
or $\left(\text{mass}\right)^{2n^{2}\Delta}$.

As a next step, the conformal transformation $z=\exp\frac{2\pi}{L}w$
maps the cylinder to the complex plane parametrized by the dimensionless
complex coordinates $z$ and $\bar{z}$. Under this transformation,
the vertex operators $V_{\pm1}$ transform as conformal primary fields
of left/right weights $(\Delta,\Delta)$ \cite{BPZ}:

\begin{equation}
V_{\pm1}^{\mathrm{pl}}(z,\bar{z})\left(|z|\,\frac{2\pi}{L}\right)^{2\Delta}=V_{\pm1}^{\mathrm{cyl}}(w,\bar{w})\,,\label{MapPlaneCyl}
\end{equation}
where unlike the vertex operators $V^{\mathrm{cyl}}$ defined on the
cylinder, vertex operators $V^{\mathrm{\mathrm{pl}}}$ defined on
the plane are dimensionless. Hence \eqref{MapPlaneCyl} allows to
express \eqref{eq:pcft_Hamiltonian} as\footnote{The $\vartheta$ integral runs over the unit circle as $z=e^{i\vartheta}$,
corresponding to $\tau=0$ on the cylinder.}

\begin{equation}
H_{\mathrm{PCFT}}=\frac{2\pi}{L}\left(L_{0}+\bar{L}_{0}-\frac{1}{12}\right)-\lambda\left(\frac{2\pi}{L}\right)^{2\Delta}\frac{L}{2}\int_{0}^{2\pi}\frac{\ud\vartheta}{2\pi}\left[V_{+1}^{\mathrm{pl}}(e^{i\vartheta},e^{-i\vartheta})+V_{-1}^{\mathrm{pl}}(e^{i\vartheta},e^{-i\vartheta})\right]{\color{red}\,,}\label{HPCFT}
\end{equation}
i.e. in terms of a dimensionless cosine operator $(V_{+1}^{\mathrm{pl}}+V_{-1}^{\mathrm{pl}})/2$,
which can be straightforwardly matched to the corresponding operator
$:\cos\varphi_{i}:$ in the lattice regularized approach. Note that
any two definitions of the exponential operator are related by some
multiplicative renormalisation using the Baker-Campbell-Hausdorff
formula \eqref{eq:BCH}. Operators $(V_{+1}^{\mathrm{pl}}+V_{-1}^{\mathrm{pl}})/2$
and $:\cos\varphi_{i}:$ have identical normalization since they both
have expectation value $1$ in the vacuum state of the massless free
boson defined by setting $\lambda=0$ in the PCFT and $J=0$ on the
lattice. A similar result is true for the relation between the sine
operators.

The first part of the Hamiltonian (\ref{HPCFT}) involves the generators
$L_{0}$ and $\bar{L}_{0}$ of the Virasoro algebra and is just the
free massless boson Hamiltonian in finite volume, which can be rewritten
in terms of the usual bosonic operators as

\begin{equation}
H_{\mathrm{CFT}}=\frac{2\pi}{L}\left(\pi_{0}^{2}+\sum_{k>0}a_{-k}a_{k}+\sum_{k>0}\bar{a}_{-k}\bar{a}_{k}-\frac{1}{12}\right)\:,\label{HCFT}
\end{equation}
with

\begin{equation}
\begin{split}[\phi_{0},\pi_{0}]=i\qquad & [a_{k},a_{l}]=k\delta_{k+l}\\{}
[\bar{a}_{k},\bar{a}_{l}]=k\delta_{k+l}\,,
\end{split}
\end{equation}
where $\phi_{0}$ and $\pi_{0}$ are the zero mode of the canonical
field and its conjugate momentum. The operators $a_{k}$ and $\bar{a}_{k}$
correspond to right and left oscillator modes creating/annihilating
particles with momentum $p=\pm2\pi|k|/L$.

The Hilbert space $\mathcal{H}$ is composed of Fock modules $\mathcal{F}_{n}$,
built upon Fock vacua $|n\rangle=V_{n}(z=0)|0\rangle$ using the oscillator
modes, and its basis is given as

\begin{equation}
a_{-k_{1}}...\,a_{-k_{r}}\bar{a}_{-p_{1}}...\,\bar{a}_{-p_{l}}|n\rangle:\,n\in\mathbb{Z}\,,\,r,\,l\in\mathbb{N}\:,k_{i}\,,\,p_{j}\in\frac{2\pi}{L}\mathbb{N}^{+}\,,\label{Hilbert}
\end{equation}
which are eigenstates of $H_{\mathrm{CFT}}$ with energy 
\begin{equation}
E=\frac{2\pi}{L}\left(\frac{(n\beta)^{2}}{4\pi}+\sum_{i=1}^{r}k_{i}+\sum_{j=1}^{l}p_{j}-\frac{1}{12}\right)\,.
\end{equation}
The ground state of the conformal field theory is the Fock vacuum
with $n=0$, i.e. $|0\rangle$.

Note that the PCFT Hamiltonian is obtained by setting (Euclidean)
time to $\tau=0$. In the subsequent calculations we use this Hamiltonian
for time evolution, which means that we use a Schrödinger picture
in which operators are time-independent and states evolve under the
full Hamiltonian in contrast with the usual conformal field theory
picture, where the operators are evolved by the conformal Hamiltonian
\eqref{HCFT}. The two pictures are physically equivalent via a similarity
transformation by the operator 
\[
e^{-\tau H_{\text{CFT}}}\,.
\]
Finally, it is useful to measure all quantities in units of the first
breather mass $m_{1}$, and define a dimensionless volume variable
as 
\begin{equation}
l=m_{1}L\,.\label{eq:l}
\end{equation}
This normalization implies that distance is measured in units of the
Compton wave length corresponding to the first breather: 
\begin{equation}
\ell_{1}=\dfrac{\hbar}{m_{1}c}\,.\label{eq:Compton}
\end{equation}

\section{How TWA and TCSA work \label{sec:TWA_TCSA}}

This section treats the methods used to simulate the time evolution:
first the truncated Wigner approximation in Sec. \ref{subsec:TWA},
then the truncated conformal space approach Sec. \ref{subsec:TCSA},
and then describes the parameter matching between the two approaches
in Sec. \ref{subsec:ParameterMatching}.

\subsection{Truncated Wigner approximation \label{subsec:TWA}}

The truncated Wigner approximation (TWA) is a powerful semiclassical
method. It is constructed through a systematic expansion of the Keldysh
path integral \cite{polkovnikov,polkovnikov2} and is well suited
for calculating out-of-equilibrium expectation values and correlations.
In this section we present the TWA formulas for the lattice Hamiltonian
\eqref{eq:HLat}, while for the sake of completeness we review the
detailed derivation of the method in App. \ref{app:twa} following
Refs.~\cite{polkovnikov,polkovnikov2}.

Our main purpose here is the calculation of the out-of-equilibrium
expectation value $\langle\hat{\mathcal{O}}\rangle(t)$ of an arbitrary
operator $\hat{\mathcal{O}}$, for an initial state given by the density
matrix $\hat{\rho}_{0}$, and a time evolution governed by the Hamiltonian
\eqref{eq:HLat}. First we introduce the notations $|\varphi\rangle_{j}$
and $|n\rangle_{j}$ for the eigenstates of the operators $\hat{\varphi}_{j}$
and $\hat{n}_{j}$, respectively. These eigenstates satisfy the completeness
relations 
\begin{equation}
{\rm \mathbb{I}_{j}}=\int_{-\pi}^{\pi}\dfrac{\ud\varphi}{2\pi}|\varphi\rangle_{j\,j}\langle\varphi|=\sum_{n\in\mathbb{Z}}|n\rangle_{j\,j}\langle n|\label{eq:complete}
\end{equation}
at any site $j$, while their overlap is given by 
\begin{equation}
_{j}\langle\varphi|n\rangle_{j}=e^{i\varphi n}\,.\label{eq:overlap}
\end{equation}
Relying on the observation that the phase never winds by the full
period $2\pi$ in our simulations, in the following we neglect the
$2\pi$ periodicity of the phase, which would ensure that the particle
number operator takes integer values. In this approximation both $\varphi$
and $n$ become continuous variables, and the completeness relation
\eqref{eq:complete} is replaced by 
\begin{equation}
{\rm \mathbb{I}_{j}}=\int_{-\infty}^{\infty}\dfrac{\ud\varphi}{2\pi}|\varphi\rangle_{j\,j}\langle\varphi|=\int_{-\infty}^{\infty}\ud n\,|n\rangle_{j\,j}\langle n|.\label{eq:complete2}
\end{equation}
Below we will use a more compact vector notation 
\begin{equation}
\underline{\varphi}=\lbrace\varphi_{j}\,|j=1,...,N_{s}\rbrace
\end{equation}
for the full set of eigenvalues, with analogous notation for the eigenvalues
of the particle number operators $\hat{n}_{j}$.

In the TWA we express the expectation value $\langle\hat{\mathcal{O}}\rangle(t)$
in terms of the Wigner function of the initial state, 
\begin{align}
W(\underline{\varphi},\underline{n})=\dfrac{1}{(2\pi)^{2N_{s}}}\int\ud\underline{\varphi}^{\prime}\,\langle\underline{\varphi}+\underline{\varphi}^{\prime}/2|\,\hat{\rho}_{0}\,|\underline{\varphi}-\underline{\varphi}^{\prime}/2\rangle\,e^{-i\underline{\varphi}^{\prime}\underline{n}},\label{eq:W}
\end{align}
with $\hat{\rho}_{0}$ denoting the density matrix at $t=0$, and
in terms of the Wigner transform of the operator $\hat{\mathcal{O}}$,
\begin{align}
O_{W}(\underline{\varphi},\underline{n})=\dfrac{1}{(2\pi)^{N_{s}}}\int\ud\underline{\varphi}^{\prime}\,\langle\underline{\varphi}-\underline{\varphi}^{\prime}/2|\,\hat{\mathcal{O}}\,|\underline{\varphi}+\underline{\varphi}^{\prime}/2\rangle\,e^{i\underline{\varphi}^{\prime}\underline{n}}.\label{eq:Ow}
\end{align}
The resulting TWA expression can be written in a compact form as 
\begin{equation}
\langle\hat{\mathcal{O}}\rangle_{\mathrm{TW}}(t)=\int\int\ud\underline{\varphi}\,\ud\underline{n}_{0}\,W(\underline{\varphi}_{0},\underline{n}_{0})\,O_{W}(\underline{\varphi}(t),\underline{n}(t))\,,\label{eq:TW}
\end{equation}
where the components of the trajectories $\underline{\varphi}(t^{\prime})$
and $\underline{n}(t^{\prime})$ are determined by the following classical
equations of motion, 
\begin{align}
 & \partial_{t}n_{j}=-\dfrac{Kc}{\pi a}(\varphi_{j+1}+\varphi_{j-1}-2\varphi_{j})-\dfrac{2J\rho_{0}a}{\hbar}\sin\varphi_{j}\,,\nonumber \\
 & \partial_{t}\varphi_{j}=\dfrac{c\pi}{Ka}n_{j}\,,\label{eq:EOM}
\end{align}
solved for initial conditions $\{\underline{\varphi}_{0},\underline{n}_{0}\}$
(see App. \ref{app:leading} for more details).

Using the TWA result \eqref{eq:TW}, the time evolution of observables
can be evaluated by the following procedure. We generate random initial
conditions $\{\underline{\varphi}_{0},\underline{n}_{0}\}$, drawn
from the distribution given by the initial Wigner function $W(\underline{\varphi}_{0},\underline{n}_{0})$,
and we obtain the classical trajectories from \eqref{eq:EOM}. Substituting
the fields $\underline{\varphi}(t)$ and $\underline{n}(t)$ into
the Wigner transform $O_{W}$ gives a single realization of the observable.
The TWA expectation value \eqref{eq:TW} of the observable at time
$t$ is then evaluated as the average of $O_{W}(\underline{\varphi}(t),\underline{n}(t))$
over realizations corresponding to a large number of different initial
conditions.

As already mentioned at the beginning of this section, the TWA arises
as the leading order contribution of a systematic expansion of the
Keldysh path integral in terms of quantum fields (see App. \ref{app:leading}).
In the App. \ref{app:correction} we also examine the next term of
this expansion, and compare the resulting quantum correction to the
TWA result, Eq.~\eqref{eq:TW}.

\subsection{Truncated conformal space approach \label{subsec:TCSA}}

The truncated conformal space approach (TCSA) is an efficient numerical
method to study perturbed conformal field theories, originally introduced
in \cite{TCSAYurovZamo}. The main idea is to consider the theory
of interest in a finite volume $L$ resulting in a discrete spectrum
of the unperturbed CFT, which can be truncated to a finite subspace
by introducing an upper energy cut-off parameter $e_{\mathrm{cut}}$.
For many perturbations of CFTs it is possible to calculate exact matrix
elements of the perturbing field and various operators in the truncated
Hilbert space. Therefore, computing the spectrum of the perturbed
theory and other physical quantities reduces to manipulations with
finite dimensional matrices.

For the sine-Gordon TCSA \cite{c1TCSA} the starting point is the
Hamiltonian (\ref{HPCFT}) of a compactified free massless boson in
finite volume $L$, perturbed by a relevant cosine operator with the
Hilbert space in finite volume spanned by the basis (\ref{Hilbert}).
Using the simplest truncation scheme described above, the truncated
space is given by 
\begin{equation}
\mathcal{H}_{\mathrm{TCSA}}(e_{\mathrm{cut}})=\text{span}\left\{ a_{-k_{1}}...\,a_{-k_{r}}\bar{a}_{-p_{1}}...\,\bar{a}_{-p_{l}}|n\rangle:\:\frac{(n\beta)^{2}}{4\pi}+\sum_{i=1}^{r}k_{i}+\sum_{j=1}^{l}p_{j}-\frac{1}{12}\leq e_{\mathrm{cut}}\right\} 
\end{equation}
which is the scheme commonly employed in the literature. To keep our
notations compact, we return to the conformal field theoretical convention
$\hbar=1$ and $c=1$ in this section.

In our investigations of the time evolution it is more convenient
to use a different truncation scheme with two parameters. The number
of Fock modules is fixed by requiring $|n|\leq n_{\mathrm{cut}}$,
but within each module we also apply a module independent energy cut-off
$e_{\mathrm{cut}}$. This prescription leads to the following truncated
Hilbert space: 
\begin{equation}
\mathcal{H}_{\mathrm{TCSA}}(e_{\mathrm{cut}},n_{\mathrm{cut}})=\text{span}\left\{ a_{-k_{1}}...\,a_{-k_{r}}\bar{a}_{-p_{1}}...\,\bar{a}_{-p_{l}}|n\rangle:\:\sum_{i=1}^{r}k_{i}+\sum_{j=1}^{l}p_{j}\leq e_{\mathrm{cut}}\;,|n|\leq n_{\mathrm{cut}}\right\} \,.\label{eq:our_TCS}
\end{equation}
The effect of the in-module energy cut-off $e_{\mathrm{cut}}$ can
partially be eliminated using ideas inspired by the renormalisation
group as discussed in App. \ref{sec:TCSARG} in more detail. There
are no analogous methods to compensate the effect of the other truncation
parameter $n_{\text{cut}}$ , so we chose an alternative approach.
For the time evolution of a state, one can choose a suitable fixed
value of $n_{\text{cut}}$ by requiring that the norm of the component
of the time evolved state falls inside the extremal Fock modules remain
small for the time period considered in the simulation. This condition
can be easily checked during the numerical time evolution. Although
the calculation of expectation values of operators with respect to
a time evolved state requires some additional care, choosing $n_{\text{cut}}$
using the above self-consistent monitoring yields a controllable approximation.

Matrix elements of the vertex operators $V_{m}$ can easily be computed
in the conformal basis using the mode expansion of the canonical field
$\phi$ on the cylinder: 
\begin{equation}
\phi(x,t)=\phi_{0}+\frac{4\pi}{L}\pi_{0}t+i\sum_{k\text{\ensuremath{\neq}0}}\frac{1}{k}\left[a_{k}\exp\left(i\frac{2\pi}{L}k(x-t)\right)+\bar{a}_{k}\exp\left(-i\frac{2\pi}{L}k(x+t)\right)\right]\,.
\end{equation}
It is straightforward to show that the matrix elements of the vertex
operators 
\begin{equation}
\langle n'|a_{k'_{1}}...\,a_{k'_{r'}}\bar{a}_{p'_{1}}...\,\bar{a}_{p'_{l'}}V_{m}a_{-k_{1}}...\,a_{-k_{r}}\bar{a}_{-p_{1}}...\,\bar{a}_{-p_{l}}|n\rangle
\end{equation}
are independent on the Fock module index of the states apart from
a selection rule $\delta_{n',n+m}$. Therefore, using the Fock decomposition
of the free boson Hilbert space 
\[
\mathcal{H}=\bigoplus_{_{n}}\mathcal{F}_{n}\,,
\]
the Hamiltonian of sine-Gordon model has a simple modular structure
which can be represented as a tri-diagonal block matrix, where the
entries correspond to operators acting either within each block (the
conformal part $H_{0}$) or between neighboring Fock modules (the
blocks $\mathcal{V}_{\pm1}$ from the vertex operators $V_{\pm1}$):
\begin{align}
H_{\mathrm{TCSA}}= & \left(\begin{array}{ccccccc}
\ddots & \ddots & \ddots\\
 & \mathcal{V}_{1} & H_{0}^{(n+1)} & \mathcal{V}_{-1}\\
 &  & \mathcal{V}_{1} & H_{0}^{(n)} & \mathcal{V}_{-1}\\
 &  &  & \mathcal{V}_{1} & H_{0}^{(n-1)} & \mathcal{V}_{-1}\\
 &  &  &  & \ddots & \ddots & \ddots
\end{array}\right)\,.\label{eq:tcsa_block_form}
\end{align}
This matrix is finite dimensional when restricted to the space \eqref{eq:our_TCS},
and its numerical diagonalisation yields an approximation of the energy
levels and corresponding eigenstates of the model.

It was demonstrated in \cite{QuenchTCSA} that TCSA is also an efficient
tool to compute the time evolution by directly constructing the action
of the truncated evolution operator $e^{-itH_{\text{TCSA}}}$; in
the present work this was achieved by using pre-programmed algorithms
to compute the action of a matrix exponential on a vector.

An important limitation of both the TCSA and the TWA is an upper limit
on the evolution time due to the finite volume used in the calculation.
Namely, for a calculation in a volume $L=l/m_{1}$ the time evolution
only follows the infinite size system for as long as the excitations
do not have time to get around the volume to affect the observable
considered. For the one-point observables considered in this work
this limit is $m_{1}t\leq l$ (in units with $c=\hbar=1$). For times
longer than this upper limit one can see the effects of the periodic
boundary condition. This sets the upper time limit for simulation
results presented in the next Section.

On the other hand, for TCSA another important limitation arises preventing
the reach of the weak coupling regime of the sine-Gordon theory, which
manifests in the need for a large number of relevant Fock modules,
resulting in a large Hilbert space. The difficulties of TCSA in the
regime of weak interactions may seem counter-intuitive at first since
the weaker the interaction, the smaller the fluctuations in the ground
state of the uncoupled system. However, Eq.~\eqref{eq:m1L} implies
that for weaker interactions, a larger initial number of particles
is required to compensate the smaller value of $J$ to keep the value
of the dimensionless volume $l=L/\ell_{1}$ fixed, while fluctuations
of the particle number asymmetry only decrease in proportion to the
total particle number, but not in absolute magnitude. Of course $l$
can also be decreased to reduce the fluctuations; however, smaller
volume not only decreases the upper time limit accessible by the TCSA
evolution, but for values $l\lesssim10$ (i.e. $L\lesssim10\ell_{1}$
with the Compton length $\ell_{1}=\hbar/(m_{1}c)$ setting the correlation
length $\xi_{corr}\approx\ell_{1}$), one expects strong finite size
corrections to the infinite volume sine-Gordon dynamics.

\subsection{Parameter matching\label{subsec:ParameterMatching}}

The sine-Gordon theory has a natural correlation length $\xi_{\mathrm{corr}}$
which can be identified as the Compton wave length $\ell_{1}$ given
in Eq. \eqref{eq:Compton}. To compare results in the lattice and
perturbed conformal field theory formulation, $\xi_{\mathrm{corr}}$
must be larger than the lattice spacing. Since we also compare dynamical
quantities, it is also necessary that the lattice regularized dispersion
relation \eqref{eq:LatticeDispRel} be linear in the energy range
influencing the dynamics. Given these conditions, the only remaining
task is to express the dimensionless volume parameter $l$ of the
PCFT, Eq. \eqref{eq:l}, in terms of the lattice parameters, since
the only other parameter of the QFT $\beta$ is already expressed
in terms of the Luttinger parameter $K$ in \eqref{eq:betaK}. In
this section we restore $\hbar$ and $c$ explicitly to obtain the
relations in terms of the physical units used in the experiments.
It is then convenient to introduce the dimensionless coupling $\kappa$
by rewriting \eqref{eq:massgap} in the form $\lambda/(\hbar c)=\kappa\,\ell_{1}^{2\Delta-2}$.
Using \eqref{eq:Hlat} and \eqref{HPCFT} we obtain the relation 
\begin{equation}
\kappa\,\ell_{1}^{2\Delta-2}\left(\frac{2\pi}{L}\right)^{2\Delta}=\frac{2J\rho_{0}}{\hbar c}\mathcal{N}.
\end{equation}
From these relations the mass of the first breather in the field theory
is expressed in terms of the lattice quantities as 
\begin{equation}
m_{1}=\dfrac{\hbar}{c}\left(\frac{2J\rho_{0}}{\kappa\hbar c}\mathcal{N}\left(\frac{L}{2\pi}\right)^{2\Delta}\right)^{\frac{1}{2-2\Delta}}\:,\label{eq:PhysMass}
\end{equation}
and the dimensionless volume turns out to be 
\begin{equation}
l=L/\ell_{1}=\left(\frac{JLN\mathcal{N}}{\kappa\hbar c(2\pi)^{2\Delta}}\right)^{\frac{1}{2-2\Delta}}\,.\label{eq:m1L}
\end{equation}
To compare dynamical quantities we recall that in lattice simulations
it is customary to measure time in units of the (bare/unrenormalized)
Josephson time $T_{J}=1/f_{J}$ with 
\begin{equation}
f_{J}=\sqrt{\frac{J}{h}\frac{c}{2KL}N}\,.\label{f_J}
\end{equation}
Here $f_{J}$ arises as the oscillation frequency in the single mode
approximation of the lattice Hamiltonian \eqref{eq:HLat}, with homogeneous
phase and particle number difference $\varphi_{j}\equiv\varphi_{0}$
and $n_{j}\equiv n_{0}$, 
\begin{equation}
H_{{\rm sing.m.}}=\dfrac{\hbar c\pi}{2KL}n_{0}^{2}-2J\rho_{0}L\cos\varphi_{0}
\end{equation}
within the harmonic approximation $\cos\varphi_{0}\approx1-\varphi_{0}^{2}/2$.
Note that $H_{{\rm sing.m.}}$ coincides with the pendulum considered
in the introduction with the identification $U=\hbar c\pi/(2KL)$.
On the other hand, in the QFT the convenient dimensionless variable
is $\nu_{1}t$, with the frequency $\nu_{1}$ associated with the
breather mass, 
\begin{equation}
\nu_{1}=\dfrac{m_{1}c^{2}}{h}.\label{eq:nu1}
\end{equation}
It is then easy to calculate the relation between the dimensionless
times $f_{J}\,t_{\mathrm{Lat}}$ and $\nu_{1}t_{QFT}$ and eventually
between $f_{J}$ and $\nu_{1}$ from \eqref{eq:m1L}, \eqref{f_J}
and \eqref{eq:nu1} yielding 
\begin{equation}
\begin{split}f_{J}=\frac{\nu_{1}}{\chi}\:.\end{split}
\end{equation}
Here 
\begin{equation}
{\normalcolor \chi=\left(\frac{l}{2\pi}\right)^{\Delta}\frac{1}{\beta}\sqrt{\frac{\mathcal{N}}{\kappa}}\,,}
\end{equation}
in which $\chi$ is expressed in terms of the QFT quantities $\beta$
and $l=L/\ell_{1}$ , and the number of lattice sites $N_{s}$, but
can also be easily recast in terms of the parameters of the lattice
Hamiltonian \eqref{eq:Hlat}.

\section{Simulations}

\label{sec:TimeEvol}

In this section, we focus on two different initial states, and study
their time evolutions using the TWA and TCSA methods.{} In order
to test the methods both in and out of the regime of classical self-trapping
of the simplified pendulum model \eqref{eq:pendulum}, we consider
initial states with finite and zero particle number imbalance. For
technical reasons, TCSA can not treat initial states intersecting
the separatrix, since a state having finite weight in both phases
quickly spreads into a large number a Fock modules. We note that the
validity of TWA is also questionable in the vicinity of the phase
boundary, because here the trajectories are very sensitive to small
perturbations, which is expected to result in a large quantum correction
to the TWA.

First we consider two identical condensates with zero particle number
imbalance in Sec. \ref{sec:TimeEvolI}; here the relevant trajectories
are not self-trapped. We then impose a large enough initial particle
number difference in Sec. \ref{sec:Time-evolution-II} so that the
trajectory lies deep in the self-trapped phase.

\subsection{Two independent and identical condensates in their ground state\label{sec:TimeEvolI}}

First we consider an initial state with two identical condensates
in their ground states and well-defined atom numbers $N/2$ on each
side. In principle, one could reach this state by first cooling the
atoms in the presence of a high barrier, and then coupling them by
decreasing the barrier height to establish Josephson tunneling. In
practice, a similar state can be implemented experimentally by first
raising adiabatically the barrier between the condensates and then
waiting until they decohere. In the latter case, however, the initial
state would display large atom number fluctuations on each side.

In this setup, the phases of the two condensates are initially uncorrelated,
and therefore $\langle\cos\varphi(x)\rangle=0$ at any point $x$
at time $t=0$. Tunneling, however, leads to a build-up of phase correlations,
and gives rise to a non-zero expectation value, $\langle\cos\varphi(x)\rangle\ne0$.
This phenomenon is called phase locking. Notice that while the global
(average) phases may become almost perfectly correlated, the value
of $\langle\cos\varphi\rangle$ is always reduced by quantum fluctuations
and fluctuations due to the finite energy density of excitations after
the quench. Therefore, even for strong phase locking, $\langle\cos\varphi\rangle\lesssim1$.
In this initial state, the difference $N_{L}-N_{R}$ between left
and right particle numbers vanishes at $t=0$, and its expectation
value also remains zero at all times. Notice that \textendash{} due
to the periodic boundary conditions \textendash{} one-point functions
of local operators are position independent.

The initial state above can be implemented easily with both methods.
Within TCSA, the initial state corresponds simply to the ground state
of the unperturbed free boson CFT, while within TWA, it is described
by the Wigner function 
\begin{equation}
W=\dfrac{\theta(\varphi_{0}+\pi)\,\theta(\pi-\varphi_{0})}{2\pi}\,\delta_{n_{0},0}\,\prod_{k>0}\dfrac{4}{\pi^{2}}\exp\left(-\sigma_{k}^{2}\,\varphi_{k}\,\varphi_{-k}-\dfrac{4\,n_{k}\,n_{-k}}{\sigma_{k}^{2}}\right)\,,\label{eq:Win}
\end{equation}
with $\varphi_{0}=\sum_{j=1}^{N_{s}}\varphi_{j}/N_{s}$ the global
phase difference, $n_{0}=\sum_{j=1}^{N_{s}}n_{j}$ the difference
of atom numbers, and $n_{k\ne0}$ and $\varphi_{k\ne0}$ the standard
Fourier coefficients, 
\begin{align}
n_{k\ne0}=\dfrac{1}{\sqrt{N_{s}}}\sum_{j=1}^{N_{s}}e^{-ikja}\,n_{j} & ,\quad\quad\varphi_{k\ne0}=\dfrac{1}{\sqrt{N_{s}}}\sum_{j=1}^{N_{s}}e^{-ikja}\,\varphi_{j}\;.
\end{align}
The particle number difference $n_{0}$ takes on the well defined
value $n_{0}=0$, while the global relative phase $\varphi_{0}$ is
completely random in the interval $(-\pi,\pi)$. The variance $\sigma_{k}^{2}$
is determined by zero-point fluctuations, and is given by 
\begin{equation}
\sigma_{k}^{2}=\dfrac{4K}{\pi}\sin\dfrac{ka}{2}\approx\dfrac{2Kka}{\pi},
\end{equation}
with the last approximation valid in the regime of linear spectrum,
where $\varepsilon_{k}\approx\hbar ck$.

\begin{figure}[H]
\begin{centering}
\subfloat[$\langle\cos\varphi\rangle$ and $\mathcal{N}\langle:\cos\beta\phi:\rangle^{\text{pl}}$]{\begin{centering}
\includegraphics[width=0.45\columnwidth]{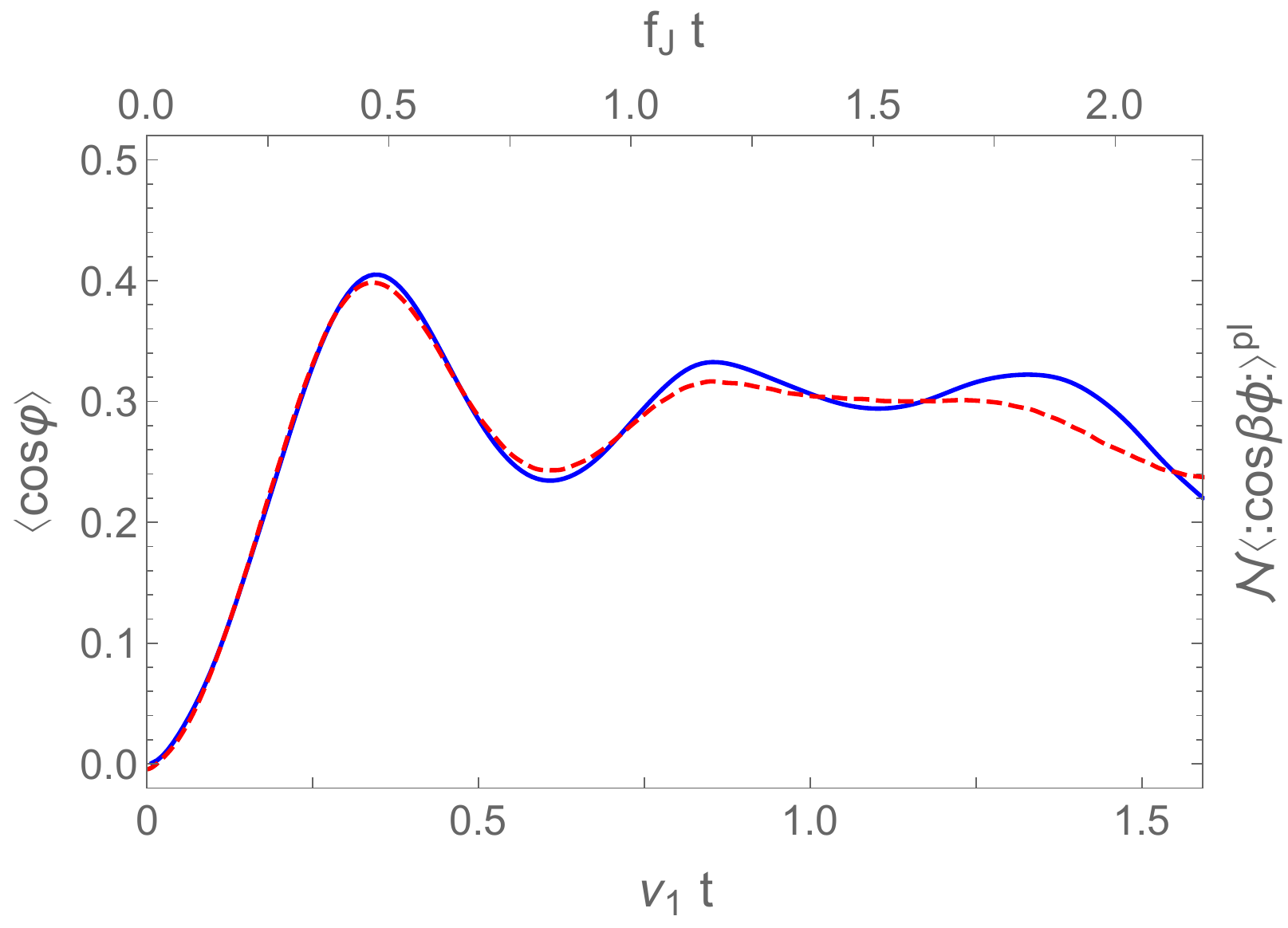} 
\par\end{centering}
}\subfloat[Standard deviation of $(N_{R}-N_{L})/2$]{\begin{centering}
\includegraphics[width=0.45\columnwidth]{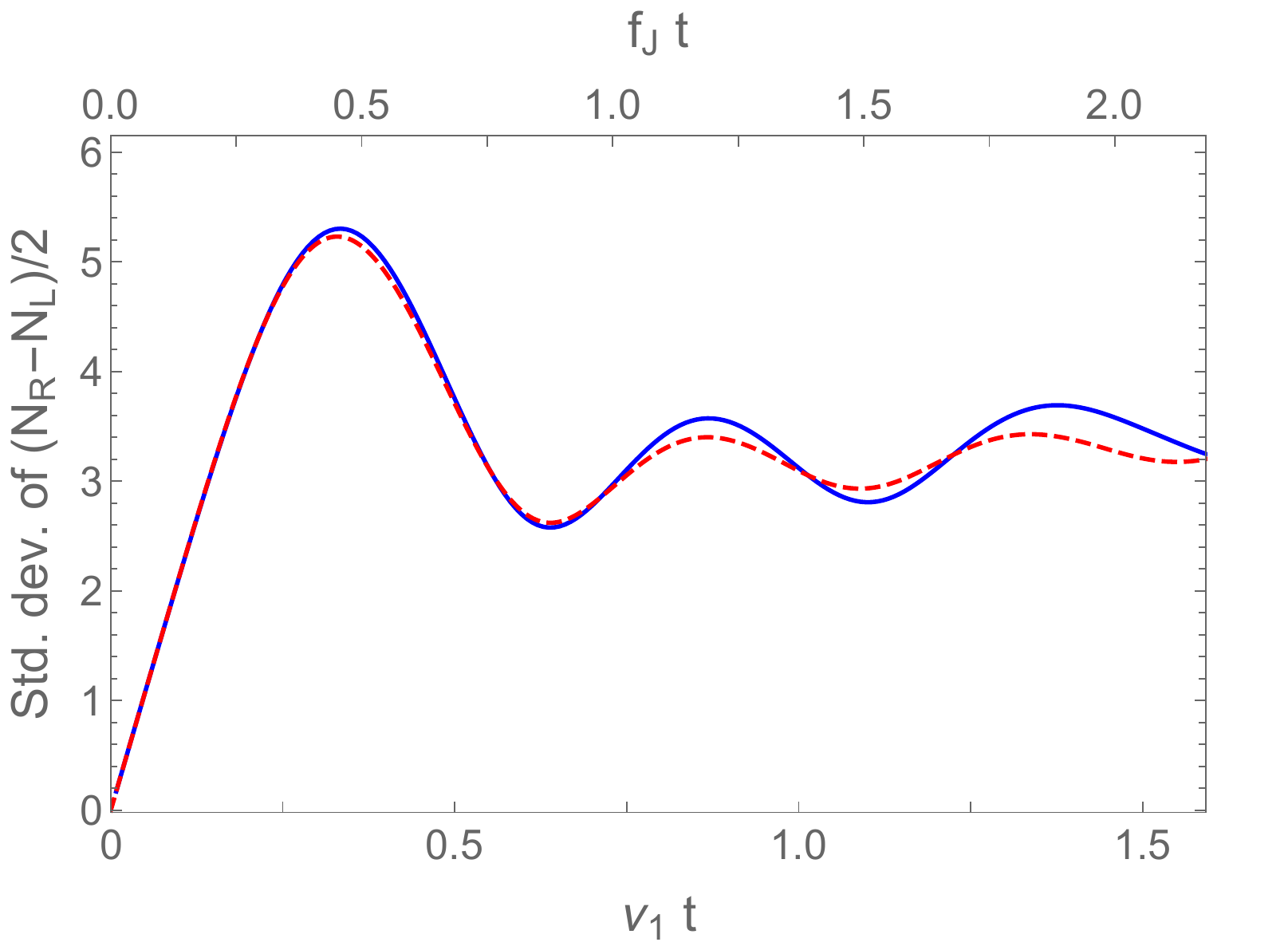} 
\par\end{centering}
}
\par\end{centering}
\caption{\label{fig:cos_and_std} Time dependent expectation value of (a) $\cos\varphi$
and $\mathcal{N}:\cos\beta\phi:^{\text{pl}}$ and (b) the standard
deviation of half of the particle number difference. Continuous blue
curves correspond to extrapolated TCSA data, while dashed red lines
to TWA. For TWA we have used $K=1.56$, $L=14.86\;\mu{\rm m}$, $N=400$,
$c=2800\;\mu m/s$, $J/h=7{\rm Hz}$ and $N_{s}=60$, corresponding
to a Josephson frequency $f_{J}=410.8$Hz. The parameters of TCSA
are $\beta=1.42$, $\nu_{1}=299.8$Hz, and the dimensionless length
$l=10$. Time evolution is measured in terms of the bare Josephson
frequency $f_{J}$ (TWA) and the renormalized Josephson frequency,
corresponding to the frequency associated with the first breather,
i.e. $\nu_{1}$ (within TCSA). Here and in all subsequent figures
the upper index ``pl'' indicates the PCFT expectation value computed
on the $(z,\bar{z})$ complex plane as specified in Sec. \ref{subsec:PertCFT}.}
\end{figure}

Fig.~\ref{fig:cos_and_std}.a displays the time evolution of $\cos\varphi$
as computed by TWA, and the corresponding quantity $\mathcal{N}\langle:\cos\beta\phi:\rangle^{\text{pl}}$
computed by TCSA, with the upper index ``pl'' referring to the PCFT
expectation value calculated on the $(z,\bar{z})$ complex plane,
as discussed in Sec. \ref{subsec:PertCFT}. Note that the expectation
value of the cosine operator calculated on the plane can be easily
expressed by that of defined on the cylinder as $\langle:\cos\beta\phi:\rangle^{\text{pl}}=\left(\frac{L}{2\pi}\right)^{2\Delta}\langle:\cos\beta\phi:\rangle^{\text{cyl}}$
according to \eqref{MapPlaneCyl}. The time evolution of the standard
deviation of the asymmetry between the number of atoms in the right
and left condensates 
\begin{equation}
\frac{\hat{N}_{R}-\hat{N}_{L}}{2}=\sum_{j=1}^{N_{s}}\hat{n}_{j}
\end{equation}
is presented in Fig.~\ref{fig:cos_and_std}.b. Within TCSA, this
variable corresponds to the quantum number $n$ labelling the Fock
modules $\mathcal{F}_{n}$.\footnote{This is also apparent from the block-diagonal form of the Hamiltonian
\eqref{eq:tcsa_block_form} since the blocks $\mathcal{V}_{\pm1}$
change $n$ by $\pm1$, while according to the bozonization relations
\eqref{eq:phase_rep} they correspond to the tunneling of an atom
from the left to the right and vice versa, respectively, described
by the two terms in Eq. \eqref{HMicroscopic}.}

The TWA results are plotted against the unrenormalized Josephson frequency,
$f_{J}$, Eq.~\eqref{f_J}, while TCSA data are presented in terms
of the frequency associated with the first breather $\nu_{1}$, Eq.~\eqref{eq:nu1},
corresponding to a renormalized Josephson frequency. Notice that the
time needs only a $\sim30\%$ rescaling, signaling that even for the
strong interactions corresponding to the system analyzed in Fig. \ref{fig:cos_and_std},
renormalisation effects are sizable but not crucial for the experimentally
relevant system sizes.

The TCSA curves were obtained by implementing a renormalisation group-based
extrapolation, outlined in App. \ref{sec:TCSARG}, using the raw results
with energy cut-offs $e_{cut}=12,14,16$ and $18$ as input. The other
TCSA truncation parameter $n_{cut}$ was set to 11, which ensured
that the norm of the component of the time evolved state in the extremal
Fock modules $\mathcal{F}_{11}$ and $\mathcal{F}_{-11}$ remained
smaller than $10^{-3}$ for the time range considered.

The numerical results obtained by the two methods show good agreement
within the accessible time frame. Estimating the errors of the two
numerical methods is a rather involved task, which deserves some attention.
For the TCSA, one can get an idea about the order of magnitude of
the remaining truncation errors by investigating the cut-off dependence
of extrapolated curves. Since we have only order-of-magnitude estimates,
we chose not to indicate them directly in Fig. \ref{fig:cos_and_std},
but they remain rather small on the scale of the plots.

For the TWA, errors turn out to be much less controlled: we have estimated
quantum corrections by examining the next terms in the semiclassical
expansion of the Keldysh action and found that these become sizable
in a rather short time (see App. \ref{app:correction}). On the other
hand, the very good agreement with TCSA suggests that the \emph{actual}
error is much smaller than our estimate, and can be crudely estimated
by the deviation between the two simulation results. We thus do not
have at present a good way to control the accuracy of TWA, which nevertheless
performs in this case surprisingly well.

The results presented in this subsection appear to disagree with those
of recent experiments, which reported a rapid build-up of partial
phase coherence between the condensates, followed by a somewhat slower
relaxation to a phase-locked steady state with $\langle\cos\varphi\rangle\approx1$
\cite{SchmiedmayerPhase}. Instead, here we find that $\langle\cos\varphi\rangle$
quickly approaches a stationary value considerably lower than 1 for
the sine-Gordon model. Even though the experiments of the Schmiedmayer
group were performed on a weakly interacting system, while our results
in Fig. \ref{fig:cos_and_std} have been obtained for a Luttinger
parameter $K=1.56$, corresponding to strong atom-atom interactions,
we would observe a similar stationary value for weaker interactions.
We return to a more detailed discussion of this issue in the conclusions.

\subsection{Two condensates in their ground state with a particle number difference}

\label{sec:Time-evolution-II}

\begin{figure}[H]
\begin{centering}
\subfloat[$\langle\cos\varphi\rangle$ and $\mathcal{N}\langle:\cos\beta\phi:\rangle^{\text{pl}}$]{\begin{centering}
\includegraphics[width=0.45\columnwidth]{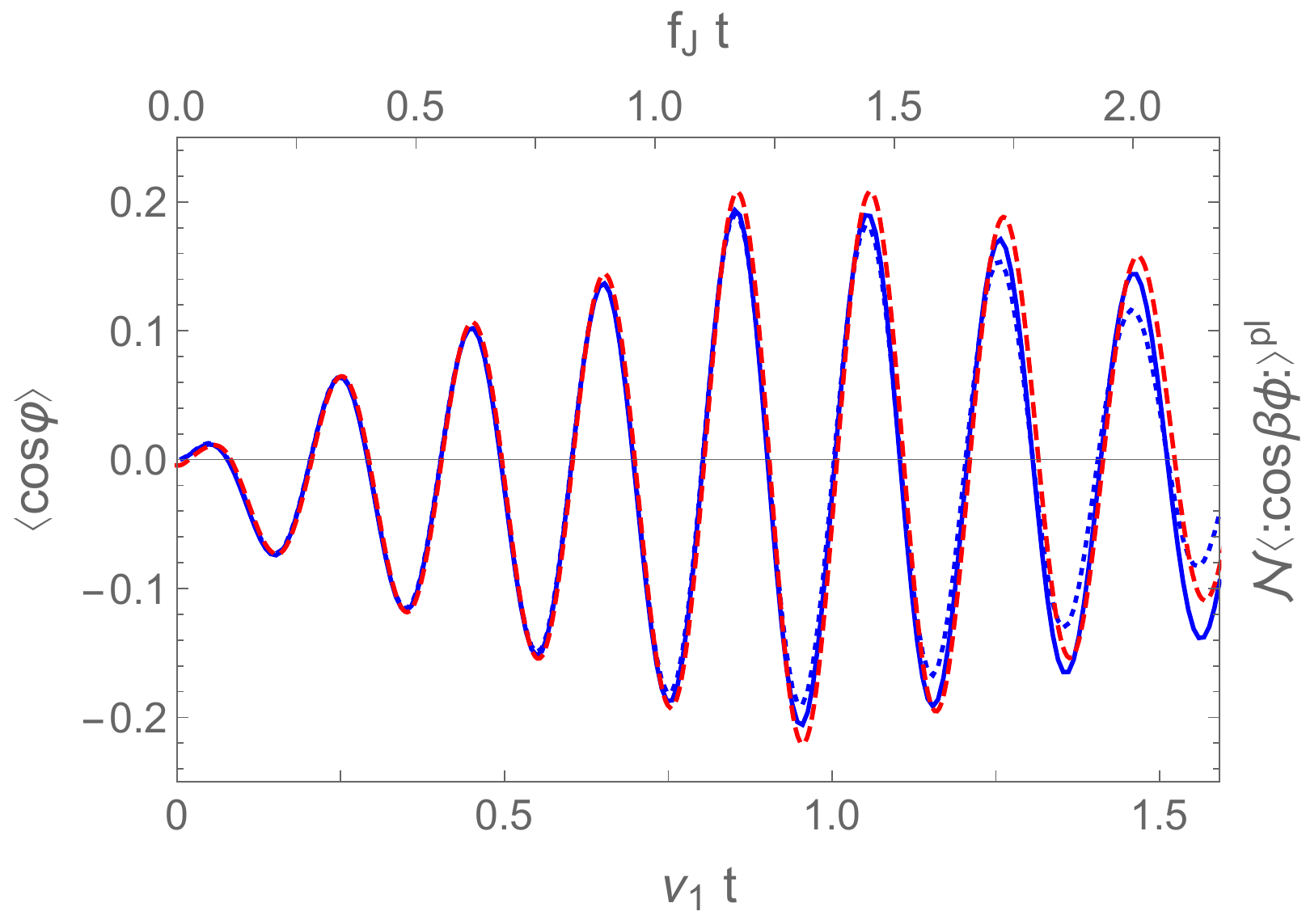} 
\par\end{centering}
}\subfloat[$\langle\sin\varphi\rangle$ and $\mathcal{N}\langle:\sin\beta\phi:\rangle^{\text{pl}}$]{\begin{centering}
\includegraphics[width=0.45\textwidth]{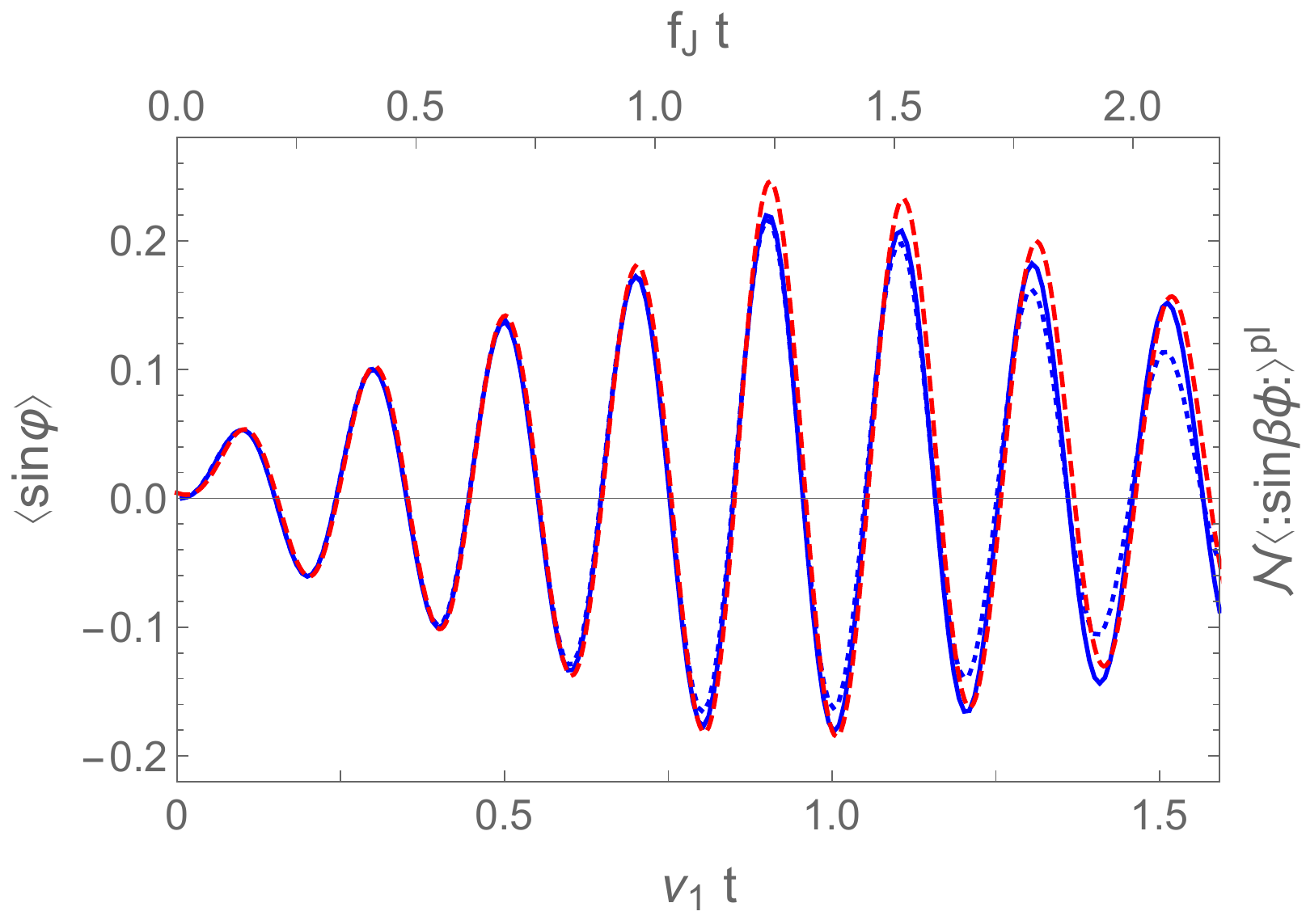} 
\par\end{centering}
}
\par\end{centering}
\caption{\textbf{\label{fig:cosine_and_sine} }Comparing results for (a) $\cos\varphi$
and $\mathcal{N}:\cos\beta\phi:^{\text{pl}}$, and (b) $\sin\varphi$
and $\mathcal{N}:\sin\beta\phi:^{\text{pl}}$, obtained from TWA and
TCSA, respectively. Blue continuous curve corresponds to TCSA with
$e_{\text{cut}}=18$, blue dotted curve to the extrapolated TCSA data
and red dashed curve to TWA, for an initial particle number imbalance
$\left(N_{R}-N_{L}\right)/2=25$. The parameters of TWA are $K=1.56$,
$L=14.86\;\mu{\rm m}$, $N=400$, $c=2800\;\mu m/s$, $J/h=7{\rm Hz}$
and $N_{s}=60$, corresponding to a Josephson frequency $f_{J}=410.8$Hz.
For TCSA we used $\beta=1.42$, $\nu_{1}=299.8$Hz, and the dimensionless
length $l=10$. Time evolution is measured in terms of the bare (TWA)
and renormalized (TCSA) Josephson frequency, $f_{J}$ and $\nu_{1}$,
respectively.}
\end{figure}

\begin{figure}[H]
\begin{centering}
\subfloat[$\langle(N_{R}-N_{L})/2\rangle$ ]{\begin{centering}
\includegraphics[width=0.45\columnwidth]{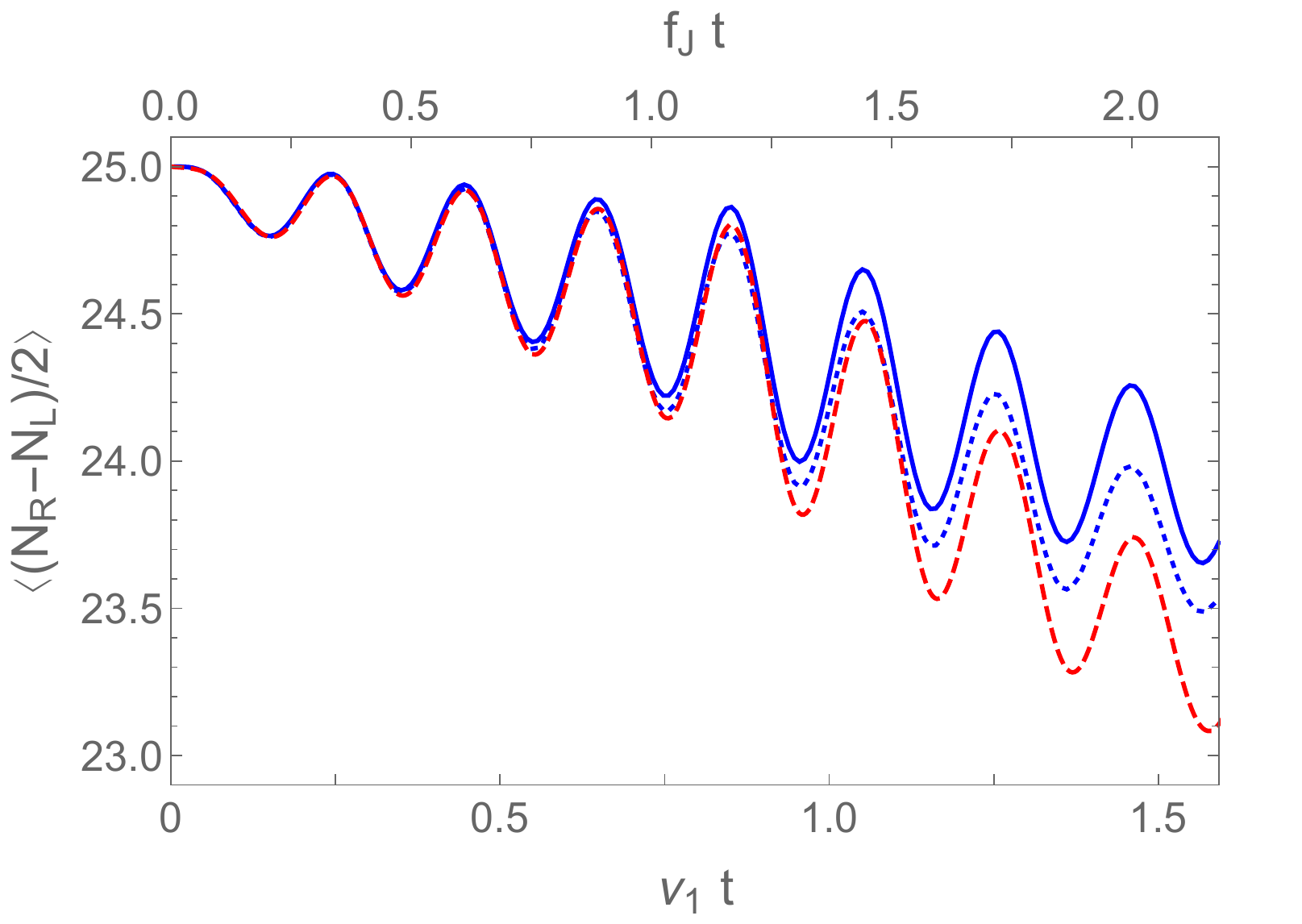} 
\par\end{centering}
}\subfloat[Standard deviation of $(N_{R}-N_{L})/2$ ]{\begin{centering}
\includegraphics[width=0.45\textwidth]{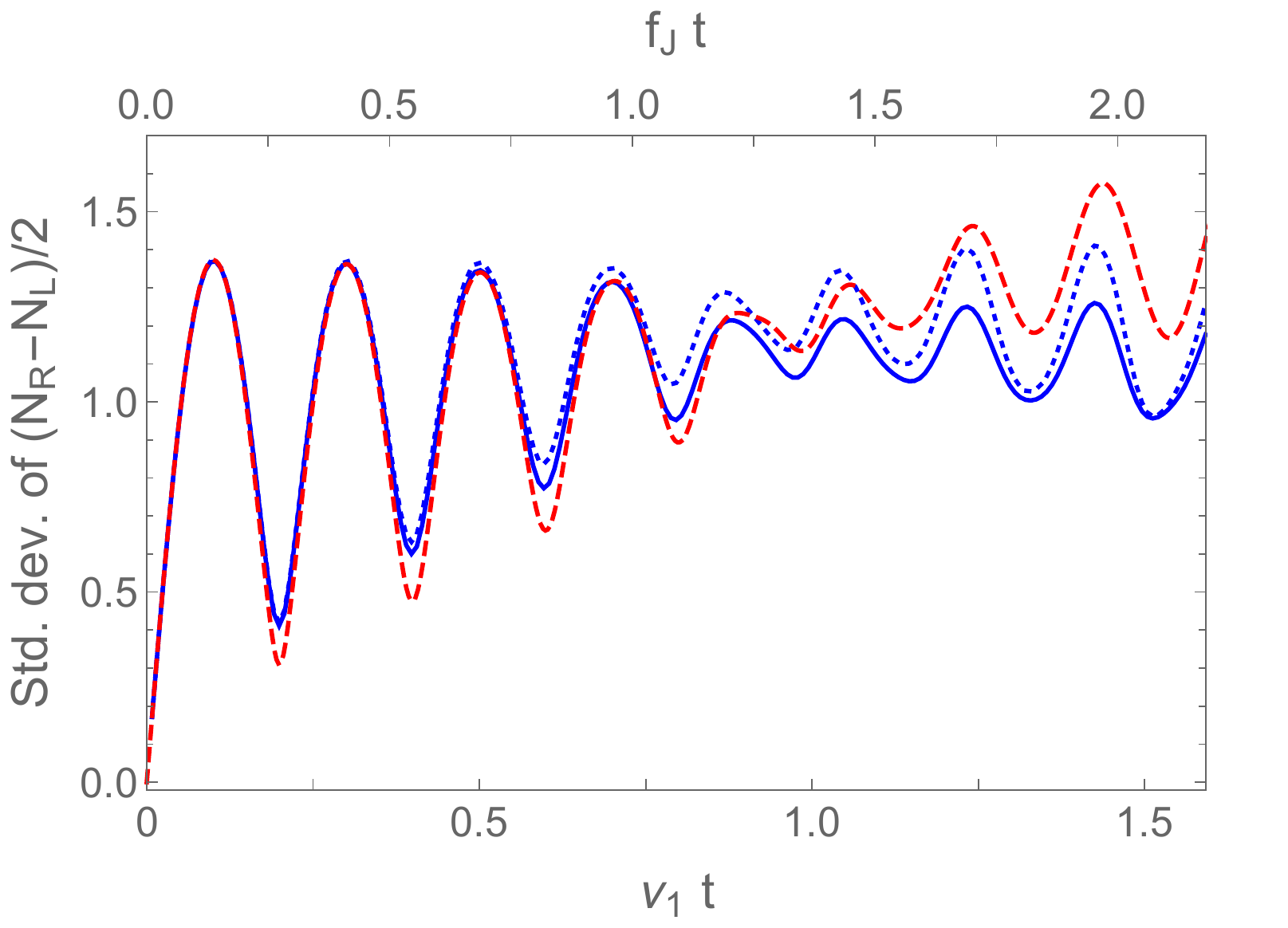} 
\par\end{centering}
}
\par\end{centering}
\caption{\textbf{\label{fig:asym_and_its_std} }Time dependent expectation
value of half of the particle number difference (a) and its standard
deviation (b), for an initial particle number imbalance $\left(N_{R}-N_{L}\right)/2=25$,
with blue continuous, blue dotted and red dashed curves corresponding
to TCSA with $e_{\text{cut}}=18$, to the extrapolated TCSA data and
to TWA, respectively. The parameters of TWA are $K=1.56$, $L=14.86\;\mu{\rm m}$,
$N=400$, $c=2800\;\mu m/s$, $J/h=7{\rm Hz}$ and $N_{s}=60$, corresponding
to a Josephson frequency $f_{J}=410.8$Hz. For TCSA we used $\beta=1.42$,
$\nu_{1}=299.8$Hz, and the dimensionless length $l=10$. Time is
measured in units of the bare (TWA) and renormalized (TCSA) Josephson
frequency, $f_{J}$ and $\nu_{1}$, respectively.}
\end{figure}

We now turn to the investigation of the regime of classical self-trapping.
To this end we discuss the situation where the two condensates are
prepared in their ground states, but with a large enough initial particle
number asymmetry, $\left(N_{R}-N_{L}\right)/2=N_{0}$, such that the
initial state is far enough from the separatrix as shown in Fig. \ref{fig:pendulum}.
(In practice, we monitor the number of relevant Fock modules in TCSA,
and we ensure that the state does not leak into the non-trapped phase
during the time evolution.) In TCSA, this state is the ground state
$|N_{0}\rangle$ of the Fock-module $\mathcal{F}_{N_{0}}$, while
in the TWA it corresponds to the Wigner function \eqref{eq:Win},
but with the factor $\delta_{n_{0},0}$ replaced by $\delta_{n_{0},N_{0}}$.
The evolution of the expectation values of the cosine and the sine
of the relative phase are presented in Fig. \ref{fig:cosine_and_sine},
while the expectation value and standard deviation of the particle
number difference are shown in Fig. \ref{fig:asym_and_its_std}. Note
that unlike in the previous setting, the sine of the phase and the
particle number asymmetry are both non-vanishing due to the asymmetry
in the initial state.

For the one point functions $\cos\varphi$ and $\sin\varphi$ the
match between the TCSA and TWA is excellent when considering the blue
continuous curve (corresponding to TCSA with $e_{\text{cut}}=18$)
and the red dashed curve (TWA). However, the TCSA extrapolation drives
away the curves for slightly larger times. In contrast, for the standard
deviation of $(N_{R}-N_{L})/2$ and especially for its average, the
extrapolation results in better agreement. This is due to the initial
state being highly excited, resulting in a larger extrapolation error
for the available values of the cut-off (see App. \ref{sec:TCSARG}),
which is also manifested in the quality of the individual extrapolation
fits. Therefore it is not possible to decide a priori whether the
extrapolated curve or the one with the highest energy cut-off is closer
to the correct PCFT result. However, this uncertainty is negligible
for shorter times, therefore in this regime the sine-Gordon time evolution
is captured correctly both by the TWA and the TCSA.

For the TCSA extrapolation in energy, runs with cut-off values $e_{\text{cut}}=12,\,14,\,16$
and $18$ were used. In addition, for all the TCSA simulations $n_{\text{cut}}=11$
was chosen, which means that Fock-modules $\mathcal{F}_{n}$ from
the range $n_{0}-n_{\text{cut}}\leq n\leq n_{0}\text{+\ensuremath{n_{\text{cut}}}}$
(i.e. $\mathcal{F}_{14},\ldots,\mathcal{F}_{36}$) were included in
the truncated Hilbert space. With this choice, the square of the norm
of the time evolved state remained less than $10^{-7}$ in the extremal
Fock modules $\mathcal{F}_{14}$ and $\mathcal{F}_{36}$ during the
entire time evolution.

All one-point functions displayed in Figs. \ref{fig:cosine_and_sine}
and \ref{fig:asym_and_its_std} show pronounced oscillations with
a period $T$ much smaller than the bare Josephson time $T_{J}$,
\begin{equation}
T/T_{J}\approx0.3.
\end{equation}
We can understand these faster oscillations by considering the classical
trajectories of the pendulum \eqref{eq:pendulum}. The period of the
trajectory of energy $E$ is given by 
\begin{equation}
T_{E}=\oint\dfrac{\ud\varphi_{0}}{\dot{\varphi}_{0}}=\hbar\oint\dfrac{\ud\varphi_{0}}{2Un_{0}}=\dfrac{\hbar}{2\sqrt{U}}\oint\dfrac{\ud\varphi_{0}}{\sqrt{E+NJ\cos\varphi_{0}}},
\end{equation}
with the integral running along the trajectory. With the parametrization
of Eq. \eqref{eq:pendulum}, the bare Josephson time, the period corresponding
to the lowest energy $E_{0}=-JN$, is given by 
\begin{equation}
T_{J}=\dfrac{h}{\sqrt{2UNJ}}.
\end{equation}
Comparing $T_{J}$ to the period of a self-trapped trajectory, we
arrive at 
\begin{equation}
\dfrac{T_{E}}{T_{J}}=\dfrac{\sqrt{2}}{\pi\sqrt{1+E/(NJ)}}\left\lbrace F\left(\dfrac{\pi}{2}\middle|\dfrac{2}{1+E/(NJ)}\right)-F\left(-\dfrac{\pi}{2}\middle|\dfrac{2}{1+E/(NJ)}\right)\right\rbrace ,
\end{equation}
with $F$ denoting the elliptic integral of the first kind. By evaluating
this expression for the trajectory touched by our initial state at
$\varphi_{0}=0$ (see Fig. \ref{fig:pendulum}), corresponding to
energy 
\begin{equation}
E=\dfrac{\hbar c\pi}{2KL}N_{0}^{2}-JN,
\end{equation}
we find $T_{E}/T_{J}=0.3$, in excellent agreement with the numerics.

Moreover, the expectation values plotted in Fig. \ref{fig:cosine_and_sine}
show a pronounced beating. This effect can be qualitatively understood
by noting that the classical trajectories intersected by the initial
state fall in a small frequency window, and the dominant contribution
to the dynamics comes from the vicinity of the trajectories touched
at $\varphi_{0}=0$ and $\varphi=\pi$ (see Fig. \ref{fig:pendulum}).
The period of the beating can be estimated from the frequency shift
between these two trajectories, leading to 
\begin{equation}
1/T_{b}=1/T_{E_{1}}-1/T_{E_{2}}
\end{equation}
with 
\begin{equation}
E_{1,2}=\dfrac{\hbar c\pi}{2KL}N_{0}^{2}\pm JN.
\end{equation}
These considerations result in the estimate $T_{b}/T_{J}\approx1.8$,
which is by a factor of 1.5 smaller than the beating period observed
in the numerics. This discrepancy most likely originates from the
presence of $k\text{\ensuremath{\neq}0}$ modes, oscillating with
a frequency slightly shifted compared to the frequency of the zero
mode, and strongly renormalizing the period of beating.

We note that the $k\neq0$ modes also have a pronounced effect on
the dynamics of the particle number difference $(N_{L}-N_{R})/2$,
plotted in Fig. \ref{fig:asym_and_its_std}. On the top of oscillations,
$(N_{L}-N_{R})/2$ decreases gradually, since the large excitation
energy stored in the zero mode in the initial state is transferred
to the $k\neq0$ modes in the course of the time evolution.

\section{Conclusions \label{sec:Conclusions}}

In this work, we studied real-time out-of-equilibrium time evolution
in the quantum sine-Gordon model by comparing the truncated Wigner
approximation (TWA) and the truncated conformal space approach (TCSA).
Quantum quenches in the sine-Gordon model have received considerable
interest recently, especially in light of experiments involving cold
atomic gases such as coupled quasi-one-dimensional bosonic condensates
\cite{Schmiedmayer,SchmiedmayerPhase}, in which the effective description
of the dynamics is thought to be provided by sine-Gordon theory. Therefore,
besides establishing connection between the parametrization of the
two numerical methods together with linking these parameters to the
experimental ones, we also concentrated on studying quench protocols
relevant to experimental investigations.

Whereas the applied numerical approaches are relatively easy to implement,
estimating their systematic error and therefore their applicability
is a difficult, and for the TWA an essentially unsolved problem. In
the case of the TCSA, the RG-based cut-off extrapolation provides
a useful improvement as well as giving an idea of the magnitude of
truncation errors from the quality of the cut-off extrapolation fits.
The TCSA method is expected to face difficulties for highly excited
states which was indeed found to be the case.

For the sine-Gordon TWA, in contrast to TWA based on coherent state
representation, no natural small parameter emerges to control the
expansion of the Keldysh path integral, and hence neglecting quantum
corrections is not guaranteed to capture essential physics. In fact
we found that the evaluation of the next quantum correction seriously
overestimates the error, and therefore the approximation is largely
uncontrolled. Therefore comparison with results obtained by the alternative
method of TCSA has a high value, since a good agreement is strong
evidence for the reliability of the results.

We have studied two very different initial conditions. In the first
case, we assumed two independent condensates, which are then connected
by a Josephson tunneling term at time $t=0$. A quick initial rise
of the expectation value of $\langle\cos\varphi\rangle$ has been
observed with both methods, which, followed by a few oscillations,
leveled off at a value $\langle\cos\varphi\rangle<1$. The other initial
state we have studied assumed a large particle number difference at
time $t=0$. In this second case the system was in the trapped phase
and simultaneous beating effects and oscillations appeared.

For the particular parameters of the sine-Gordon model and the quench
protocols discussed in the paper, excellent agreement was found between
results for time evolved quantities obtained by the two methods for
the first few oscillations up to times of order $t/T_{J}=2$ with
$T_{J}$ denoting the Josephson time in the model (i.e., the oscillation
period within a harmonic single mode approximation for the coupled
condensates). In the language of conformal field theory, this time
scale corresponds to $c^{2}m_{1}t/\hbar=10$ , where $m_{1}$ is the
mass gap, emerging from the first breather in the attractive regime
studied here. This demonstrates the time evolution of expectation
values are correctly captured by both methods, at least for initial
states and the particular parameter range considered here.

In microtraps, the time evolution of the overall phase difference
and particle number difference is often described in terms of a simple
pendulum model \cite{SchmiedmayerPhase}. Just as ordinary pendulums,
according to this simple model, the condensate exhibits two characteristically
different behaviors, indeed observed experimentally; for small particle
number and phase differences it displays Josephson oscillations, while
for larger particle number differences a self-trapped motion appears.
Here we have tested TWA and TCSA both in the non-trapped and in the
self-trapped regimes, by considering two important \textendash{} and
also experimentally relevant \textendash{} initial states, with two
decoupled condensates prepared in their ground states. In the first
(symmetrical) case both condensates contained the same number of particles,
whereas in the second (asymmetrical) case the difference in their
particle numbers was large enough to enter the regime of classical
self-trapping\footnote{Using the language of conformal field theory, these states are eigenstates
of the free massless bosonic field theory whose perturbation with
$:\cos\beta\phi:$ results in the sine-Gordon model. The first case
corresponds to the vacuum of the free massless theory, while the second
is a similar Gaussian state with a non-zero eigenvalue with respect
to the zero mode of the canonical conjugate momentum field.}. For the symmetrical initial state, we calculated the expectation
value of the cosine of the phase of the two condensates and the standard
deviation of the particle number difference. In the asymmetrical case
these quantities were supplemented by the average of the sine of the
phase and the average of the particle number difference itself, which
vanish for the first starting condition, but are non-trivial for the
second one.

Recent experiments demonstrate that two Josephson-coupled one dimensional
condensates show a rapid relaxation to a phase-locked steady state
\cite{SchmiedmayerPhase} with $\text{\ensuremath{\langle\cos\varphi\rangle}=1}$.
This behavior is very robust against initial conditions and physical
parameters, and stands opposed to previous theoretical results \cite{Torre,Fioni1,Fioni2},
as well as to our own findings. In the symmetrical case we found that
the phase oscillates yielding values of $\text{\ensuremath{\langle\cos\varphi\rangle}}$
significantly smaller than one, and the curves are consistent with
a slow relaxation to $\text{\ensuremath{\langle\cos\varphi\rangle}<1}$,
consistent with previous theoretical results \cite{Torre}. For the
case of non-zero initial asymmetry, rapid oscillations are found whose
period matches the prediction of the pendulum model to high precision,
while their amplitude is modulated by some lower frequency. We were
able to estimate and explain the period of this beating by considering
the frequency shift between the classical trajectories of the pendulum,
intersected by our initial state, though the resulting modulation
frequency seems to be strongly renormalized by the $k\neq0$ modes.

\begin{figure}[t!]
\begin{centering}
\subfloat[$\langle\cos\varphi\rangle$]{\begin{centering}
\includegraphics[width=0.45\columnwidth]{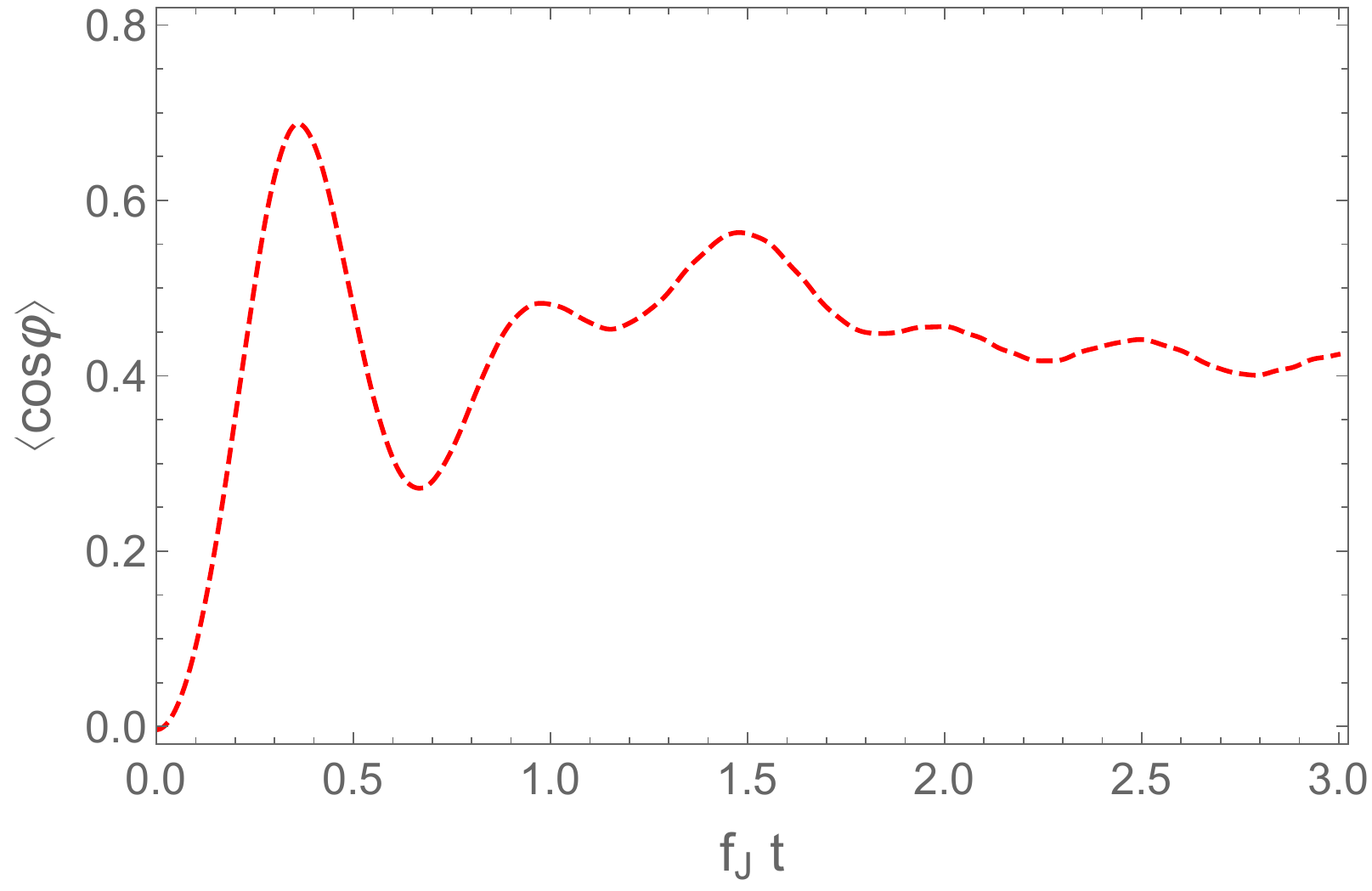} 
\par\end{centering}
}\subfloat[Standard deviation of $(N_{R}-N_{L})/2$]{\begin{centering}
\includegraphics[width=0.45\columnwidth]{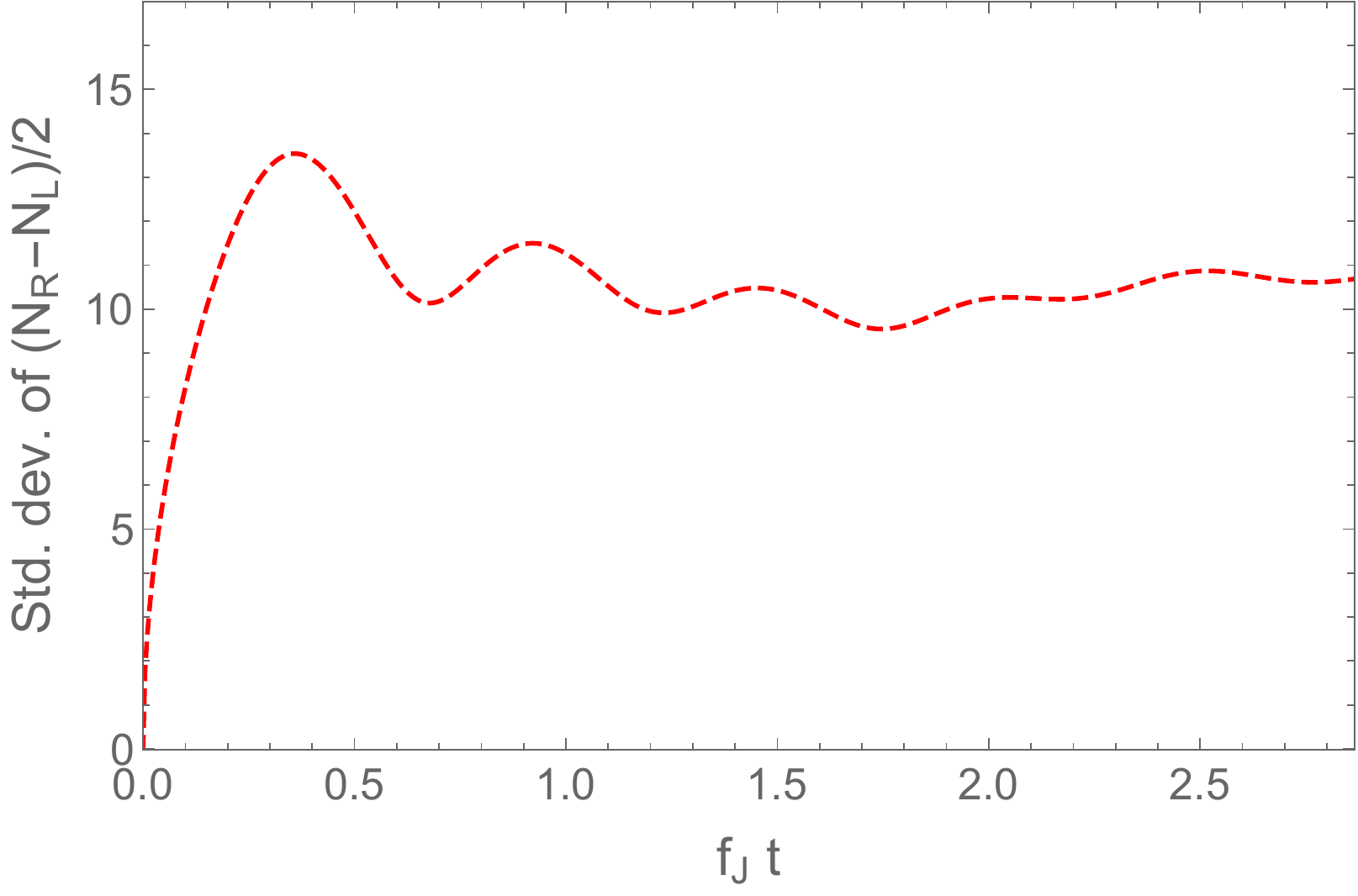} 
\par\end{centering}
}
\par\end{centering}
\caption{\label{fig:largeK}Time dependent expectation value of (a) $\cos\varphi$
and (b) the standard deviation of half of the particle number difference,
for an experimentally accessible parameter set, calculated within
TWA. We used the particle number $N=1000$, system size $L=25\mu m$,
tunnel coupling $J/h=30$Hz, and Luttinger parameter $K=27$. For
a condensate of $^{87}$Rb atoms, with atomic mass $m_{{\rm Rb}}$,
this corresponds to $c=830\mu m/s$, and typical short distance cutoff
$\xi_{h}\equiv\hbar/(m_{{\rm Rb}}c)=0.62\mu m$, consistent with the
number of lattice sites $N_{s}=L/\xi_{h}=28$. Time is measured in
terms of the bare Josephson frequency, $f_{J}=189.7$Hz.}
\end{figure}

We note that the Luttinger parameter $K$ used in our simulations
is slightly out of the reach of current experimental realizations.
To argue that our main findings, in particular, the clear absence
of fast relaxation towards a phase locked state with $\langle\cos\varphi\rangle\approx1$
holds for the weaker interaction strengths characterizing the currently
available domain of experimental parameters, we present the results
of a TWA simulation for an experimentally accessible parameter set
in Fig. \ref{fig:largeK}. As in Sec. \ref{sec:TimeEvolI}, the initial
state was the ground state of two identical condensates, and we found
that only partial coherence is built up after the recoupling, $\langle\cos\varphi\rangle\lesssim0.5$.
Unfortunately, it is difficult to use TCSA for such weak interaction
strengths, i.e. larger values of the Luttinger parameter $K$, because
the large number of relevant Fock modules results in an intractably
large Hilbert space\footnote{The seemingly counter-intuitive difficulties of TCSA in the regime
of weak interactions are discussed at the end of Section \ref{subsec:TCSA}
in more detail.}. Nevertheless, having validated the TWA by a direct comparison to
TCSA in the regime of strong interactions, the TWA results in Fig.
\ref{fig:largeK} provide strong evidence that the experimentally
observed phase locking cannot be captured by the homogeneous sine-Gordon
model considered here.

It would be interesting to have measurements directly in the range
of parameters accessible by our simulations, which could pin down
the time scale where the deviation between the sine-Gordon model and
the coupled condensate system occurs. This could help identify the
presently unknown mechanism for the experimentally found rapid phase-locking,
and is probably related to other degrees of freedom of the experimental
system. The effect of such degrees of freedom can also be studied
by including them in the TWA and/or TCSA simulations, which is an
interesting direction for future investigations.

\subsection*{Acknowledgements}

The authors are grateful to E.G. Dalla Torre, J. Schmiedmayer and
I. Mazets for useful discussions. M.K., I.L., G.T. and G.Z. are also
grateful for the hospitality of the Erwin Schrödinger Institute for
Mathematics and Physics during the thematic programme ``Quantum Paths'',
facilitating contacts with the Atomchip Group at TU Wien. This research
was supported by the National Research Development and Innovation
Office (NKFIH) under a K-2016 grant no. 119204 and an OTKA grant no.
SNN118028, and also by the BME-Nanotechnology FIKP grant of EMMI (BME
FIKP-NAT). G.T. and G.Z. acknowledge partial support by the Quantum
Technology National Excellence Program (Project No. 2017-1.2.1-NKP-2017-
00001), while M.K. was also supported by a ``Prémium\textquotedblright{}
postdoctoral grant of the Hungarian Academy of Sciences.

\appendix

\section{Mapping the coupled condensates to the sine-Gordon model\label{sec:sGMapping}\label{app:sGmap} }

In this Appendix we provide a few more technical details on the mapping
of the coupled Lieb-Liniger model given in Eqs. (\ref{eq:LiebLiniger},\ref{HMicroscopic})
to the sine-Gordon Hamiltonian \eqref{eq:H_r}.

For weak interactions, by standard bosonic commutation relations,
the phases and the density fluctuations of each condensate form conjugate
variables, $[\delta\rho_{j}(x),\varphi_{j}(y)]=-i\delta(x-y)$. Expansion
of the Lieb-Liniger Hamiltonian than yields 
\begin{equation}
H_{j}^{s}=\int\ud x\left\{ \frac{g}{2}\delta\rho_{j}^{2}(x)+\frac{\hbar^{2}\rho_{0}}{2m}[\p_{x}\fii_{j}(x)]^{2}\right\} \,,\label{eq:L1}
\end{equation}
which is the Luttinger Hamiltonian used in Refs. \cite{ExperimentalNoThermalization3,Schmiedmayer,grisins,kuhnert}.

The Luttinger Hamiltonian provides a correct description of the long
wavelength behavior even for strong interactions. Identifying now
$\delta\rho_{j}(x)$ as $\Pi_{j}(x)$, we have 
\begin{equation}
H_{j}^{s}=\frac{\hbar c}{2}\int\ud x\left\{ \frac{\pi}{K_{s}}\,\Pi_{j}^{2}(x)+\frac{K_{s}}{\pi}[\partial_{x}\varphi_{j}(x)]^{2}\right\} \,,\label{eq:Luttinger}
\end{equation}
where $[\varphi_{j}(x),\Pi_{k}(x')]=i\delta_{jk}\delta(x-x')$. The
speed of sound $c,$ and the Luttinger parameter $K_{s}$ of a single
condensate can be computed from the exact Bethe Ansatz solution of
the Lieb-Liniger model \eqref{eq:LiebLiniger}. For small and large
couplings they are given by the asymptotic formulae 
\begin{align}
K_{s} & \approx\frac{\pi}{\sqrt{\gamma}}\left(1-\frac{\sqrt{\gamma}}{2\pi}\right)^{-1/2}\approx\hbar\pi\sqrt{\frac{\rho_{0}}{mg}}\,, & c & \approx\sqrt{\frac{\rho_{0}g}{m}} & \text{for } & \gamma\lesssim10\,,\nonumber \\
K_{s} & \approx(1+4/\gamma)\,, & c & \approx\hbar\pi\rho_{0}/m & \text{for } & \gamma\gg1\,.
\end{align}
Thus for $\gamma\ll1$ Eq.~\eqref{eq:Luttinger} reduces to \eqref{eq:L1}.
Due to Galilean invariance, $cK_{s}=\hbar\rho_{0}\pi/m$ holds for
all $\gamma$.

Density fluctuations are suppressed at wavelengths smaller than the
healing length $\xi_{h}$, which also serves as a short distance cutoff.
For small $\gamma$ it is much longer than the particle-particle distance,
\begin{equation}
\xi_{h}=1/(\rho_{0}\sqrt{\gamma})=\hbar/\sqrt{mg\rho_{0}}\approx\hbar/mc\,,\label{HealingLength}
\end{equation}
while at strong coupling $\xi_{\text{h}}\approx1/\rho_{0}$ \cite{gritsev}.

The coupling between the condensates is captured by the Josephson
tunneling term, which is $2J\rho_{0}\cos(\fii_{1}-\fii_{2})$ for
small interactions, but it can be renormalized at strong interactions
\cite{gritsev}. The total Hamiltonian can therefore be rewritten
as 
\begin{equation}
H=H_{1}^{s}+H_{2}^{s}-2J\rho_{0}\int\ud x\cos\left(\varphi_{1}-\varphi_{2}\right)\,.
\end{equation}
Focusing on the evolution of the relative phase $\varphi_{1}-\varphi_{2}$,
we introduce the fields 
\begin{equation}
\begin{split}\varphi_{r}=\varphi_{1}-\varphi_{2}\:,\quad & \varphi_{t}=\frac{\varphi_{1}+\varphi_{2}}{2}\\
\Pi_{r}=\frac{\Pi_{1}-\Pi_{2}}{2}\:,\quad & \Pi_{t}=\Pi_{1}+\Pi_{2}
\end{split}
\end{equation}
which satisfy the canonical commutation relations. In our approximation,
relative and total phase degrees of freedom decouple with the relative
phase field having the Hamiltonian \eqref{eq:H_r} with the Luttinger
parameter 
\begin{equation}
K=K_{r}=K_{s}/2\:,\;.\label{eq:KrKs}
\end{equation}

\section{Relating $\cos\varphi$ and $:\cos\varphi:$\label{sec:CosFi}\label{app:normal}}

As a starting point recall the lattice Hamiltonian \eqref{eq:HLat}:
\begin{equation}
H_{Lat}=\frac{\hbar c}{2}\sum_{j=1}^{N_{s}}\left\{ \frac{\pi}{Ka}n_{j}^{2}+\frac{K}{\pi a}\left(\varphi_{j}-\varphi_{j-1}\right)^{2}\right\} -2J\rho_{0}a\sum_{j=1}^{N_{S}}\cos\varphi_{j}\,,\label{eq:HLatApp}
\end{equation}
where $a$ is the lattice spacing, $\left[\varphi_{j},n_{k}\right]=i\delta_{jk}$
with $\dot{\varphi_{i}}=\frac{c\pi}{Ka}n_{i}$ , $N_{s}=L/a$ is the
number of sites, and periodic boundary conditions are assumed. Our
goal is to express $H_{Lat}$ with the normal-ordered cosine as follows

\begin{equation}
H_{Lat}=\frac{\hbar c}{2}\sum_{j=1}^{N_{s}}\left\{ \frac{\pi}{Ka}n_{j}^{2}+\frac{K}{\pi a}\left(\varphi_{j}-\varphi_{j-1}\right)^{2}\right\} -2J\rho_{0}a\mathcal{N}\sum_{j=1}^{N_{S}}:\cos\varphi_{i}:\,.\label{eq:HLatApp2}
\end{equation}
Consider the mode expansion of the fields using the Fourier representation
\begin{equation}
\varphi_{i}=\tilde{\varphi}_{0}+\frac{1}{\sqrt{N_{s}}}\sum_{k\in\frac{2\pi}{L}\mathbb{Z},\,k\neq0}e^{ikx_{j}}\tilde{\varphi}_{k}
\end{equation}
with $x_{j}=ja$, as follows 
\begin{eqnarray}
\begin{array}{cc}
\varphi_{j}= & \frac{1}{\sqrt{N_{s}}}{\displaystyle \sum_{k\neq0}}A_{k}\left(e^{ikx_{j}-i\omega_{k}t}b_{k}+e^{-ikx_{j}+i\omega_{k}t}b_{k}^{\dagger}\right)+\varphi_{0}+\frac{\pi_{0}t}{N_{s}}\frac{c\pi}{Ka}\\
\\
n_{j}= & \frac{1}{\sqrt{N_{s}}}{\displaystyle \sum_{k\neq0}}(-i)\frac{Ka}{c\pi}\omega_{k}A_{k}\left(e^{ikx_{j}-i\omega_{k}t}b_{k}-e^{-ikx_{j}+i\omega_{k}t}b_{k}^{\dagger}\right)+\frac{\pi_{0}}{N_{s}}\,,
\end{array}\label{eq:ModeExp}
\end{eqnarray}
The zero modes $\tilde{\varphi}_{0}=\frac{1}{N_{s}}\sum_{j}\varphi_{j}$
and $\tilde{\pi}_{0}=\sum_{j}n_{j}$ satisfy 
\begin{equation}
\left[\tilde{\varphi}_{0},\tilde{\pi}_{0}\right]=\left[\varphi_{0},\pi_{0}\right]=i\,,
\end{equation}
while choosing 
\begin{equation}
A_{k}=\sqrt{\frac{c\pi}{2Ka\omega_{k}}}
\end{equation}
ensures 
\begin{equation}
\left[b_{k},b_{k'}^{\dagger}\right]=\delta_{k,k'}\,.
\end{equation}
The lattice dispersion relation stated in Eq. \eqref{eq:LatticeDispRel}
in the main text, 
\begin{equation}
\varepsilon_{k}\equiv\hbar\omega_{k}=\frac{2\hbar c}{a}\left|\sin\frac{ka}{2}\right|\,,
\end{equation}
easily follows from the $J=0$ equation of motion 
\begin{equation}
\ddot{\varphi}_{j}=\frac{c^{2}}{a^{2}}\left(\varphi_{j+1}+\varphi_{j-1}-2\varphi_{j}\right)\,.
\end{equation}

To relate $\cos\varphi$ with $:\cos\varphi:\,$, first consider the
exponential 
\begin{equation}
e^{i\varphi_{j}}=\prod_{k\neq0}\exp\left(\frac{i}{\sqrt{N_{s}}}A_{k}\left(e^{ikx_{j}-i\omega_{k}t}b_{k}+e^{-ikx_{j}+i\omega_{k}t}b_{k}^{\dagger}\right)\right)\times\exp\left(i\varphi_{0}+i\frac{\pi_{0}t}{N_{s}}\frac{c\pi}{Ka}\right)\,.
\end{equation}
This can be reordered using the Baker-Campbell-Hausdorff formula,
\begin{equation}
e^{X}e^{Y}=e^{X+Y}e^{\frac{1}{2}c}\label{eq:BCH}
\end{equation}
valid when $\left[X,Y\right]=c$ is a c-number. This results in the
relation 
\begin{align}
:e^{\text{i\ensuremath{\varphi_{j}}}}: & =\mathcal{N}e^{\text{i\ensuremath{\varphi_{j}}}}\\
\text{with } & \mathcal{N}=\prod_{k\neq0}\exp\left(\frac{c\pi}{2N_{s}a2K\omega_{k}}\right)\exp\left(i\frac{tc\pi}{2N_{s}Ka}\right).\nonumber 
\end{align}
Omitting a complex phase originating from the zero mode, we arrive
at 
\begin{equation}
\cos\varphi_{i}=\mathcal{N}:\cos\varphi_{i}:\,
\end{equation}
with the renormalisation factor written as 
\begin{align}
\mathcal{N} & =\prod_{k\neq0}\exp\left(-\frac{\pi}{8N_{s}K|\sin\frac{ka}{2}|}\right)=\prod_{n=-N_{s}/2+1(\neq0)}^{N_{s}/2}\exp\left(-\frac{\pi}{8N_{s}K|\sin\frac{\pi n}{N_{s}}|}\right)\nonumber \\
 & =\exp\left(-\frac{\pi\Delta}{N_{s}}\right)\prod_{n=1}^{N_{s}/2-1}\exp\left(-\frac{2\pi\Delta}{N_{s}\sin\frac{\pi n}{N_{s}}}\right)\,,
\end{align}
where 
\begin{equation}
\Delta=\frac{\beta^{2}}{8\pi}.\label{eq:deltaApp}
\end{equation}

\section{Cut-off dependence and extrapolation in TCSA\label{sec:TCSARG}}

\label{app:cutoff}

TCSA inevitably involves an energy cut-off $e_{\text{cut}}$ to truncate
the Hilbert space to a finite dimensional one. Therefore all quantities
computed from TCSA possess a cut-off dependence which can be addressed
using renormalisation group methods \cite{TCSARG1,TCSARG2,RGTCSAWatts,TCSARychkov,LencsesTakacs}.
Here we avoid the technical details and simply present the relevant
results for expectation values of local operators together with a
heuristic justification borrowed from \cite{TCSARGOnePoint}.

Let us denote the vacuum expectation value of a local operator in
the sine-Gordon TCSA with a cut-off parametrized as 
\begin{equation}
n=\frac{e_{\text{cut}}}{2}
\end{equation}
as $\text{\ensuremath{\langle\mathcal{O}\rangle}}^{(n)}$. It was
shown in \cite{TCSARGOnePoint} that the leading cut-off dependence
can be written as

\begin{equation}
\text{\ensuremath{\langle\mathcal{O}\rangle}}^{(n)}=\text{\ensuremath{\langle\mathcal{O}\rangle}}^{(\infty)}+\sum_{A}K_{A}n^{2\alpha_{A}-2}\left(1+O\left(\frac{1}{n}\right)\right)\,,\label{TCSAOnePointExtrap}
\end{equation}
where $\text{\ensuremath{\langle\mathcal{O}\rangle}}^{(\infty)}$
is the expectation value with the cut-off removed. Using this relation,
data points obtained for a sequence of sufficiently high $n$ can
be extrapolated numerically to obtain a precise estimate for the expectation
value $\text{\ensuremath{\langle\mathcal{O}\rangle}}^{(\infty)}$.
The exponents $\alpha_{A}$ in (\ref{TCSAOnePointExtrap}) can be
analytically determined via the operator product expansion (OPE) of
the perturbing field $V$ (i.e. the cosine potential for the sine-Gordon
theory) and the operator $\mathcal{O}$. According to the OPE, the
short distance singularity of operator products (inserted into correlation
functions) is given by

\begin{equation}
\mathcal{O}(z,\bar{z})V(w,\bar{w})\sim\sum_{A}\frac{C_{\mathcal{O}V}^{A}A(w,\bar{w})}{\left(z-w\right)^{(h_{O}+h_{V}-h_{A})}\left(\bar{z}-\bar{w}\right)^{(\bar{h}_{O}+\bar{h}_{V}-\bar{h}_{A})}}\,,
\end{equation}
where $A$ runs over a complete set of local operators, and $h$ and
$\bar{h}$ are the right and left conformal weights of the operators.
For scalar operators satisfying $h=\bar{h}$, $\alpha_{A}$ is given
by

\begin{equation}
\alpha_{A}=h_{O}+h_{V}-h_{A}\,.
\end{equation}
For a heuristic understanding of the expression (\ref{TCSAOnePointExtrap})
note that the running coupling characterizing a relevant operator
becomes small at high energy (i.e. short distance) scales. Therefore
as the dependence of expectation values on large values of the cut-off
$n$ are concerned, the corrections to the exact expectation value
$\langle T\mathcal{O}\exp\left(-\lambda\int d^{2}zV\right)\rangle$
can be replaced by ones computed from the first order perturbative
expression $-\lambda\langle T\mathcal{O}\int d^{2}zV\rangle$. The
OPE yields

\begin{equation}
\mathcal{O}\int\ud^{2}zV\sim\sum_{A}C_{A}A\,,
\end{equation}
where the dimensions of the coefficients $C_{A}$ are $[\text{energy}]^{-2+2\alpha_{A}}$.
For a large cut-off $n$, the associated energy scale $\frac{4\pi}{L}n$
is much larger then any other scale in the theory and is therefore
the only relevant energy scale. Therefore a simple scaling argument
predicts that the cut-off dependence of $\text{\ensuremath{\langle\mathcal{O}\rangle}}^{(n)}$
must be of the form $n^{2\alpha_{A}-2}$. In practical application,
it is usually enough to keep the largest one among the exponents $\alpha_{A}$
predicted by the OPE.

For the particular case of sine-Gordon model, $\alpha_{A}$ can be
calculated by using the conformal weight of the vertex operators $V_{a}$,

\begin{equation}
h_{a}=\bar{h}_{a}=a^{2}\Delta\,,
\end{equation}
whereas the conformal weight of the derivative operator $\partial\phi\bar{\partial}\phi$
is $h=1$. The fusion rules for the vertex operators $V_{a}$ and
$V_{b}$ encoding the possible operator families entering their OPE
reads

\begin{equation}
[V_{a}]\times[V_{b}]=[V_{a+b}]\,,
\end{equation}
where $[V_{a}]$ signifies the appearance of the vertex operator (primary)
and its descendants, which are obtained by multiplying the exponential
with a polynomial expression of the derivatives of $\phi$. The perturbing
cosine of the sine-Gordon model is given by the combination $V_{1}+V_{-1}$.
For the observable $\mathcal{O}=:\cos\beta\phi:=(V_{1}+V_{-1})/2$,
the fusion rules imply that the families of the $\mathcal{I}$ (i.e.
$V_{0}$) and $V_{\pm2}$ enter the relevant OPE, yielding the exponents
$n^{-2+4\Delta}$, $n^{-4+2\Delta}$ and $n^{-2-4\Delta}$ for $A=\mathcal{I},\:\partial\phi\bar{\partial}\phi$
and $:\cos2\beta\phi:$ (here $\partial\phi\bar{\partial}\phi$ appears
as the descendant field of $\mathcal{I}$). For the observable $:\sin\beta\phi:$
one obtains a similar result.

The particle number difference $N_{R}-N_{L}$ is given as a spatial
integral of the field momentum $\partial_{t}\phi=\partial\phi+\bar{\partial}\phi$,
which is a combination of operators of weights $(1,0)$ and $(0,1)$
which have non-zero spins. However, a simple application of the heuristic
scaling argument using the following OPEs (resulting from Wick's theorem
for the free massless boson $\phi$) 
\begin{align}
\partial\phi(z,\bar{z})V_{a}(w,\bar{w}) & \propto\frac{iaV_{a}(w,\bar{w})}{z-w}+\text{regular terms},\nonumber \\
\bar{\partial}\phi(z,\bar{z})V_{a}(w,\bar{w}) & \propto\frac{iaV_{a}(w,\bar{w})}{\bar{z}-\bar{w}}+\text{regular terms},
\end{align}
allows one to determine the cut-off exponent. In this case we have
a separate exponent for the singular behavior of left/right movers,
e.g. for $\partial\phi$ we have 
\begin{equation}
\alpha=h_{\partial\phi}+h_{V_{a}}-h_{V_{a}}=1\qquad\bar{\alpha}=\bar{h}_{\partial\phi}+\bar{h}_{V_{a}}-\bar{h}_{V_{a}}=0
\end{equation}
leads to the naive value $-2+\alpha+\bar{\alpha}=-1$ for the extrapolation
exponent (similarly for $\bar{\partial}\phi$ with the values of $\alpha$
and $\bar{\alpha}$ interchanged). However, note that the actual perturbing
operator is $V_{+1}+V_{-1}$, and so the leading term cancels for
the combination $\partial_{t}\phi=\partial\phi+\bar{\partial}\phi$.
The next-to-leading coefficient results from considering level $1$
descendent contributions, which leads to the cut-off exponent $-2$.
For the variance of $N_{R}-N_{L}$, it is necessary to consider the
operator product with $\partial_{t}\phi(z_{1},\bar{z}_{1})\partial_{t}\phi(z_{2},\bar{z}_{2})$;
a straightforward application of the Wick theorem to compute the free
boson OPE then results in a cut-off exponent $-2$.

We finish this section with two important comments. Firstly, although
the above discussion of the cut-off dependence of one point functions
assumed the case of vacuum expectation values, the leading order cut-off
dependence is determined by the universal OPE exponents and is therefore
the same for expectation values in excited states. The only difference
is that for the validity of the leading order cut-off extrapolation
the cut-off must also be large enough compared to the energy of the
excited state under consideration.

Secondly, further improvements can be made by adding explicit counter
terms to the TCSA Hamiltonian and also to the operator \cite{sGOverlaps,LencsesTakacs},
whose coefficients are determined by renormalisation group equations
similar to (\ref{TCSAOnePointExtrap}). However, in the context of
the present work the using the leading order expressions for such
counter-terms did not result in any notable improvement compared to
the simple extrapolation procedure sketched above, therefore they
were omitted to reduce the computational costs.

\section{Expansion of Keldysh path integral}

\label{app:twa}

In this appendix we outline the expansion of the Keldysh path integral
in terms of quantum fields, following Refs. \cite{polkovnikov,polkovnikov2}.
In App. \ref{app:leading} we derive the TWA result \eqref{eq:TW}
as the leading order of this expansion, while in App. \ref{app:correction}
we determine the next quantum correction to the TWA.

\subsection{Derivation of TWA}

\label{app:leading}

In this section we first derive \eqref{eq:TW} for a general time-dependent
Hamiltonian $\hat{H}(t)$ expressed in terms of the canonical conjugate
phase and particle number operators $\lbrace\hat{\varphi}_{i},\hat{n}_{i}\rbrace$,
and then we specialize the results to Hamiltonian \eqref{eq:HLat}.
Here we denote every operator by a hat, i.e. such as $\hat{X}$ to
distinguish them from the ordinary variables $X$ entering the path
integral.

\begin{figure}[!t]
\centering{}\includegraphics[width=0.5\columnwidth]{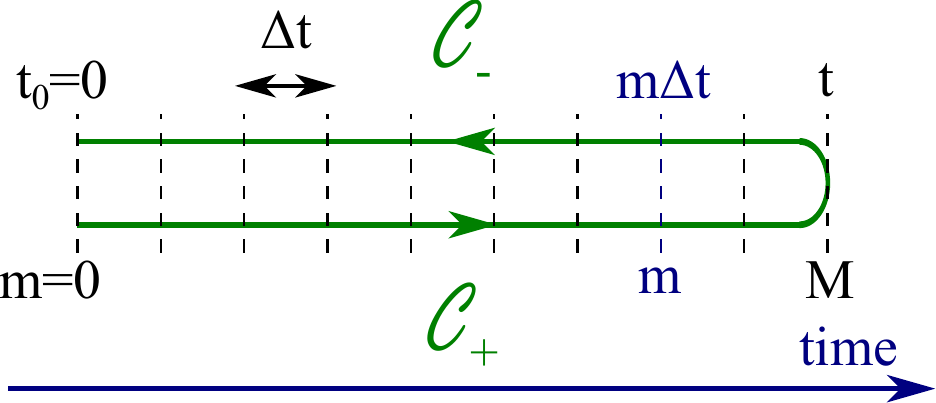} \caption{Keldysh contour. The contour of the path integral in Eq.~\eqref{eq:Oexp},
running from $t_{0}=0$ to $t$, consisting of forward ( $\mathcal{C}_{+}$)
and backward ($\mathcal{C}_{-}$) branches. All operators are ordered
along the Keldysh contour. The path integral can be evaluated by introducing
a discrete time step $\Delta t=t/M$, and inserting the completeness
relation \eqref{eq:complete2} at each time step according to Eq.~\eqref{eq:fbcontour}.}
\label{fig:contour} 
\end{figure}

For an initial density matrix $\hat{\rho}_{0}$, the out-of-equilibrium
expectation value of the operator of interest, $\hat{\mathcal{O}}$,
can be expressed as 
\begin{equation}
\langle\hat{\mathcal{O}}\rangle(t)={\rm Tr}\left(\hat{\rho}_{0}\,\mathcal{T}\,e^{-i/\hbar\int_{\mathcal{C}_{-}}\ud t^{\prime}\hat{H}(t^{\prime})}\hat{\mathcal{O}}\,e^{-i/\hbar\int_{\mathcal{C}_{+}}\ud t^{\prime}\hat{H}(t^{\prime})}\right)\,.\label{eq:Oexp}
\end{equation}
Here the operator $\mathcal{T}$ stands for the ordering along the
Keldysh contour consisting of forward and backward branches $\mathcal{C}_{+}$
and $\mathcal{C}_{-}$, with turning point $t$, depicted in Fig.
\ref{fig:contour}.

To evaluate the expectation value \eqref{eq:Oexp}, the time is discretized
in steps $\Delta t=t/M$, and the first completeness relation of \eqref{eq:complete2}
is inserted at every site $j$ after each time step on both branches
(see Fig. \ref{fig:contour}). The eigenvalue of the phase operator
$\hat{\varphi}_{j}$ at time $m\Delta t$ is denoted by $\varphi_{j,m}^{\pm}$
on the contour $\mathcal{C}_{+}/\mathcal{C}_{-}$. Similarly to the
notations of Sec. \ref{subsec:TWA}, we introduce a more compact vector
notation 
\[
\underline{\varphi}_{m}^{\pm}=\lbrace\varphi_{j,m}^{\pm}\rbrace
\]
for the full set of eigenvalues at a given time step $m$, with analogous
notations for the eigenvalues of the operators $\hat{n}_{j}$. In
the following $\hat{H}(m\Delta t)$ is abbreviated by $\hat{H}_{m}$,
also allowing for an explicit time-dependence of the Hamiltonian.

With these notations, the expectation value \eqref{eq:Oexp} can be
rewritten as 
\begin{align}
 & \langle\hat{\mathcal{O}}\rangle(t)=\int\mathcal{D}\varphi\,\langle\underline{\varphi}_{0}^{+}|\hat{\rho}_{0}|\underline{\varphi}_{0}^{-}\rangle\prod_{m=0}^{M-1}\langle\underline{\varphi}_{m}^{-}|e^{i\Delta t\hat{H}_{m+1}/\hbar}|\underline{\varphi}_{m+1}^{-}\rangle\,\langle\underline{\varphi}_{M}^{-}|\hat{\mathcal{O}}|\underline{\varphi}_{M}^{+}\rangle\prod_{m=1}^{M}\langle\underline{\varphi}_{m}^{+}|e^{-i\Delta t\hat{H}_{m-1}/\hbar}|\underline{\varphi}_{m-1}^{+}\rangle=\nonumber \\
 & \int\mathcal{D}\varphi\,\mathcal{D}n\,\langle\underline{\varphi}_{0}^{+}|\hat{\rho}_{0}|\underline{\varphi}_{0}^{-}\rangle\prod_{m=0}^{M-1}e^{i\underline{\varphi}_{m}^{-}\underline{n}_{m}^{-}}\langle\underline{n}_{m}^{-}|e^{i\Delta t\hat{H}_{m+1}/\hbar}|\underline{\varphi}_{m+1}^{-}\rangle\,\langle\underline{\varphi}_{M}^{-}|\hat{\mathcal{O}}|\underline{\varphi}_{M}^{+}\rangle\prod_{m=1}^{M}e^{i\underline{\varphi}_{m}^{+}\underline{n}_{m}^{+}}\langle\underline{n}_{m}^{+}|e^{-i\Delta t\hat{H}_{m-1}/\hbar}|\underline{\varphi}_{m-1}^{+}\rangle\,,\label{eq:fbcontour}
\end{align}
with 
\begin{equation}
\mathcal{D}\varphi=\dfrac{1}{(2\pi)^{2N_{s}(M+1)}}\prod_{m=0}^{M}d\underline{\varphi}_{m}^{+}\,d\underline{\varphi}_{m}^{-},\quad\mathcal{D}n=\prod_{m=1}^{M}d\underline{n}_{m}^{+}\,d\underline{n}_{m}^{-}\,.
\end{equation}
The second equality in \eqref{eq:fbcontour} was obtained by inserting
the second completeness relation of \eqref{eq:complete2} at each
time step, and applying \eqref{eq:overlap}. Assuming that $\hat{H}(t)$
is written in a normal ordered form, the matrix elements of the propagator
are given by 
\begin{equation}
\langle\underline{n}|e^{-i\Delta t\hat{H}_{m}/\hbar}|\underline{\varphi}\rangle=e^{-i\underline{\varphi}\underline{n}}\,e^{-i\Delta tH_{m}(\underline{n},\underline{\varphi})/\hbar}+O(\Delta t^{2})\,,
\end{equation}
where $H_{m}(\underline{n},\underline{\varphi})$ is obtained by substituting
every operator $\hat{\varphi}_{j}$ or $\hat{n}_{j}$ in $\hat{H}_{m}$
by the corresponding eigenvalue $\varphi_{j}$ and $n_{j}$, respectively.
This relation allows to express the expectation value \eqref{eq:fbcontour}
as 
\begin{align}
\langle\hat{\mathcal{O}}\rangle(t)= & \int\mathcal{D}\varphi\,\mathcal{D}n\,\langle\underline{\varphi}_{0}^{+}|\hat{\rho}_{0}|\underline{\varphi}_{0}^{-}\rangle\langle\underline{\varphi}_{M}^{-}|\hat{\mathcal{O}}|\underline{\varphi}_{M}^{+}\rangle\,e^{i\sum_{m=0}^{M-1}\underline{n}_{m}^{-}(\underline{\varphi}_{m}^{-}-\underline{\varphi}_{m+1}^{-})+i\sum_{m=1}^{M}\underline{n}_{m}^{+}(\underline{\varphi}_{m}^{+}-\underline{\varphi}_{m-1}^{+})}\nonumber \\
 & e^{i\Delta t/\hbar\sum_{m=1}^{M}\left\lbrace H_{m}(\underline{n}_{m-1}^{-},\underline{\varphi}_{m}^{-})-H_{m-1}(\underline{n}_{m}^{+},\underline{\varphi}_{m-1}^{+})\right\rbrace }\,.
\end{align}

\begin{figure}[b!]
\centering{}\includegraphics[width=0.75\columnwidth]{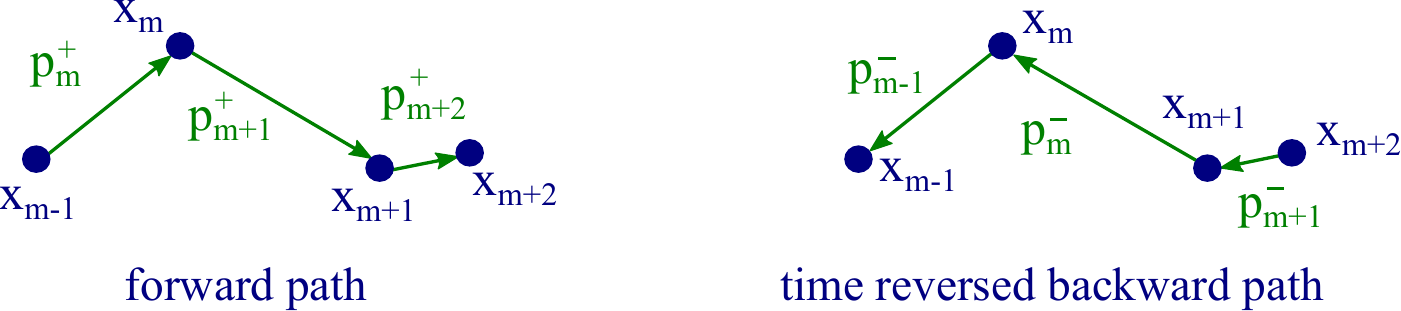}
\caption{The drawing shows time reversed pairs of forward and backward paths,
utilizing the analogy $\varphi\leftrightarrow x$ and $n\leftrightarrow p/\hbar$.
Such pairs only contribute to the classical fields in Eq.~\eqref{eq:rotation},
while they yield vanishing quantum fields.}
\label{fig:classfields} 
\end{figure}

We now introduce the classical and quantum components of the fields
$\underline{\varphi}$ and $\underline{n}$ by performing the Keldysh
rotation 
\begin{align}
 & \underline{\varphi}_{m}^{cl}=\dfrac{\underline{\varphi}_{m}^{+}+\underline{\varphi}_{m}^{-}}{2},\quad\underline{\varphi}_{m}^{q}=\underline{\varphi}_{m}^{+}-\underline{\varphi}_{m}^{-}\,,\nonumber \\
 & \underline{n}_{m}^{cl}=\dfrac{\underline{n}_{m}^{+}+\underline{n}_{m-1}^{-}}{2},\quad\underline{n}_{m}^{q}=\underline{n}_{m}^{+}-\underline{n}_{m-1}^{-}\,.\label{eq:rotation}
\end{align}
Note that the particle number on the backward path, $\underline{n}^{-}$,
is shifted by one time slice in \eqref{eq:rotation} compared to the
other fields, ensuring that the Keldysh action vanishes for purely
classical fields (i.e. when $\underline{\varphi}^{q}=\underline{n}^{q}=0$).
The origin of this index shift can be understood by noting that the
canonical conjugate variables $\varphi$ and $n$ play a role analogous
to the position $x$ and momentum $p/\hbar$ in ordinary point-particle
dynamics. In the path integral formalism the coordinates are located
on the time slices, while the momenta are assigned to the links between
them. The rotation \eqref{eq:rotation} is introduced in such a way
that the quantum fields vanish, if the backward path is the time reversal
of the forward path. As illustrated in Fig. \ref{fig:classfields},
this prescription requires to shift the momenta \textendash{} or particle
numbers \textendash{} of the backward path by one time slice, in accordance
with \eqref{eq:rotation}.

Using the new variables introduced in Eq.~\eqref{eq:rotation}, the
expectation value $\langle\hat{\mathcal{O}}\rangle(t)$ can be rewritten
as 
\begin{align}
 & \langle\hat{\mathcal{O}}\rangle(t)=\int\mathcal{D}\varphi\,\mathcal{D}n\,\langle\underline{\varphi}_{0}^{cl}+\underline{\varphi}_{0}^{q}/2|\,\hat{\rho}_{0}\,|\underline{\varphi}_{0}^{cl}-\underline{\varphi}_{0}^{q}/2\rangle\,\langle\underline{\varphi}_{M}^{cl}-\underline{\varphi}_{M}^{q}/2|\,\mathcal{\hat{\mathcal{O}}}\;|\underline{\varphi}_{M}^{cl}+\underline{\varphi}_{M}^{q}/2\rangle\times\nonumber \\
 & \quad\quad e^{i\underline{\varphi}_{M}^{q}\underline{n}_{M}^{cl}-i\underline{\varphi}_{0}^{q}\underline{n}_{1}^{cl}}\,e^{i\sum_{m=1}^{M}\underline{n}_{m}^{q}(\underline{\varphi}_{m}^{cl}-\underline{\varphi}_{m-1}^{cl})-i\sum_{m=1}^{M-1}\underline{\varphi}_{m}^{q}(\underline{n}_{m+1}^{cl}-\underline{n}_{m}^{cl})}\times\nonumber \\
 & \quad\quad e^{i\Delta t/\hbar\sum_{m=1}^{M}\left\lbrace H_{m}(\underline{n}_{m}^{cl}-\underline{n}_{m}^{q}/2,\,\underline{\varphi}_{m}^{cl}-\underline{\varphi}_{m}^{q}/2)-H_{m-1}(\underline{n}_{m}^{cl}+\underline{n}_{m}^{q}/2,\,\underline{\varphi}_{m-1}^{cl}+\underline{\varphi}_{m-1}^{q}/2)\right\rbrace }\,.\label{eq:classquant}
\end{align}
In accordance with the remark above, the exponent in the integrand
vanishes for purely classical fields, $\underline{\varphi}^{q}=\underline{n}^{q}=0$
when neglecting terms of order $\Delta t^{2}$ that disappear anyway
in the limit $\Delta t\rightarrow0$. This is a generic property ensuring
the causality structure of the Keldysh action.

Dropping boundary terms which tend to zero for $\Delta t\rightarrow0$,
the integral over $\underline{\varphi}_{0}^{q}$ can be performed,
yielding 
\begin{align}
 & \int\ud\underline{\varphi}_{0}^{q}\,\langle\underline{\varphi}_{0}^{cl}+\underline{\varphi}_{0}^{q}/2|\,\hat{\rho}_{0}\,|\underline{\varphi}_{0}^{cl}-\underline{\varphi}_{0}^{q}/2\rangle\,e^{-i\underline{\varphi}_{0}^{q}\underline{n}_{1}^{cl}}=(2\pi)^{2N_{s}}\,W(\underline{\varphi}_{0}^{cl},\underline{n}_{1}^{cl})\,.
\end{align}
Similarly, integrating over $\underline{\varphi}_{M}^{q}$ results
in 
\begin{align}
 & \int\ud\underline{\varphi}_{M}^{q}\,\langle\underline{\varphi}_{M}^{cl}-\underline{\varphi}_{M}^{q}/2|\,\hat{\mathcal{O}}\,|\underline{\varphi}_{M}^{cl}+\underline{\varphi}_{M}^{q}/2\rangle\,e^{i\underline{\varphi}_{M}^{q}\underline{n}_{M}^{cl}}=(2\pi)^{N_{s}}O_{W}(\underline{\varphi}_{M}^{cl},\underline{n}_{M}^{cl})\,.
\end{align}
The truncated Wigner approximation (TWA) is obtained by substituting
these expressions into Eq.~\eqref{eq:classquant}, and expanding
the exponent in the path integral up to first order in quantum fields,
yielding 
\begin{align}
\langle\mathcal{\hat{\mathcal{O}}} & \rangle_{TW}(t)=\int\mathcal{D}\varphi\,\mathcal{D}n\,W(\underline{\varphi}_{0}^{cl},\underline{n}_{1}^{cl})\,O_{W}(\underline{\varphi}_{M}^{cl},\underline{n}_{M}^{cl})\,e^{-i\sum_{m=1}^{M-1}\underline{\varphi}_{m}^{q}\left\lbrace \underline{n}_{m+1}^{cl}-\underline{n}_{m}^{cl}+\Delta t\,\underline{\nabla}_{\varphi}H_{m}(\underline{n}_{m}^{cl},\,\underline{\varphi}_{m}^{cl})/\hbar\right\rbrace }\times\nonumber \\
 & \qquad\qquad e^{i\sum_{m=1}^{M}\underline{n}_{m}^{q}\left\lbrace \underline{\varphi}_{m}^{cl}-\underline{\varphi}_{m-1}^{cl}-\Delta t\,\underline{\nabla}_{n}H_{m-1}(\underline{n}_{m}^{cl},\,\underline{\varphi}_{m-1}^{cl})/\hbar\right\rbrace }\nonumber \\
 & =\int\mathcal{D}\varphi^{cl}\,\mathcal{D}n^{cl}\,W(\underline{\varphi}_{0}^{cl},\underline{n}_{1}^{cl})\,O_{W}(\underline{\varphi}_{M}^{cl},\underline{n}_{M}^{cl})\,\prod_{m=1}^{M-1}\delta\left(\underline{n}_{m+1}^{cl}-\underline{n}_{m}^{cl}+\Delta t\,\underline{\nabla}_{\varphi}H_{m}(\underline{n}_{m}^{cl},\,\underline{\varphi}_{m}^{cl})/\hbar\right)\times\nonumber \\
 & \quad\prod_{m=1}^{M}\delta\left(\underline{\varphi}_{m}^{cl}-\underline{\varphi}_{m-1}^{cl}-\Delta t\,\underline{\nabla}_{n}H_{m-1}(\underline{n}_{m}^{cl},\,\underline{\varphi}_{m-1}^{cl})/\hbar\right)\,.\label{eq:TWdiscr}
\end{align}
Here $\underline{\nabla}_{\varphi}H$ and $\underline{\nabla}_{n}H$
denote the gradient of the Hamiltonian: 
\begin{equation}
\left(\underline{\nabla}_{\varphi}H\right)_{j}=\dfrac{\partial H}{\partial\varphi_{j}}\quad{\rm and}\quad\left(\underline{\nabla}_{n}H\right)_{j}=\dfrac{\partial H}{\partial n_{j}}\,,
\end{equation}
and second line of Eq.~\eqref{eq:TW} was obtained by performing
the integrals over the quantum fields using 
\begin{equation}
\int\ud\underline{\varphi}_{q}\,e^{-i\underline{\varphi}_{q}\underline{x}}=\int\ud\underline{n}_{q}\,e^{i\underline{n}_{q}\underline{x}}=(2\pi)^{N_{s}}\delta\left(\underline{x}\right)\,.
\end{equation}

Rewriting Eq.~\eqref{eq:TWdiscr} in a more compact form gives precisely
Eq.~\eqref{eq:TW}, with the trajectories $\underline{\varphi}(t^{\prime})$
and $\underline{n}(t^{\prime})$ following the classical equations
of motion, 
\begin{align}
 & \partial_{t}\underline{n}=-\underline{\nabla}_{\varphi}H(\underline{n},\,\underline{\varphi},t)/\hbar\,,\nonumber \\
 & \partial_{t}\underline{\varphi}=\underline{\nabla}_{n}H(\underline{n},\,\underline{\varphi},t)/\hbar\,,
\end{align}
solved for initial conditions $\{\underline{\varphi}_{0},\underline{n}_{0}\}$.
For the special case of Hamiltonian \eqref{eq:HLat}, these differential
equations take the form stated in Eq.~\eqref{eq:EOM}.

\subsection{Quantum corrections to TWA quantities \label{app:correction}}

In this Appendix we derive the next quantum correction to the truncated
Wigner approximation \eqref{eq:TW}, by expanding the exponent in
Eq.~\eqref{eq:classquant} up to third order in the quantum fields.
Here we restrict our attention to the specific Hamiltonian \eqref{eq:HLat},
which has a single such term of the form $(\varphi^{q})^{3}$; the
generalization for more complicated Hamiltonians is straightforward.

By expanding the integrand in \eqref{eq:classquant} as 
\begin{equation}
e^{-i\varphi_{j,m}^{q}x_{1}-i\left(\varphi_{j,m}^{q}\right)^{3}x_{2}}\approx e^{-i\varphi_{j,m}^{q}x_{1}}\left(1-i\left(\varphi_{j,m}^{q}\right)^{3}x_{2}\right)\,,
\end{equation}
and substituting \eqref{eq:W} and \eqref{eq:Ow} into Eq.~\eqref{eq:classquant},
the following correction term is obtained:

\begin{align}
\delta\langle\mathcal{O}\rangle_{1}(t) & =-i\dfrac{\Delta t}{24\,\hbar}\int\mathcal{D}\varphi\,\mathcal{D}n\,W(\underline{\varphi}_{0}^{cl},\underline{n}_{1}^{cl})\,O_{W}(\underline{\varphi}_{M}^{cl},\underline{n}_{M}^{cl})\,e^{-i\sum_{m=1}^{M-1}\underline{\varphi}_{m}^{q}\left\lbrace \underline{n}_{m+1}^{cl}-\underline{n}_{m}^{cl}+\Delta t\,\underline{\nabla}_{\varphi}H(\underline{n}_{m}^{cl},\,\underline{\varphi}_{m}^{cl})/\hbar\right\rbrace }\times\nonumber \\
 & \qquad\qquad\qquad e^{i\sum_{m=1}^{M}\underline{n}_{m}^{q}\left\lbrace \underline{\varphi}_{m}^{cl}-\underline{\varphi}_{m-1}^{cl}-\Delta t\,\underline{\nabla}_{n}H(\underline{n}_{m}^{cl},\,\underline{\varphi}_{m-1}^{cl})/\hbar\right\rbrace }\sum_{m^{\prime}=1}^{M-1}\sum_{j=1}^{Ns}\left(\varphi_{j,m^{\prime}}^{q}\right)^{3}\,\left.\dfrac{\partial^{3}H}{\partial\varphi_{j}^{3}}\right|_{\underline{\varphi}_{m^{\prime}}^{cl}}\:,
\end{align}
which can be written as 
\begin{equation}
\begin{split}\delta\langle\mathcal{O}\rangle_{1}(t)=-i\dfrac{\Delta t}{24\,\hbar}\int\mathcal{D}\varphi^{cl}\,\mathcal{D}n^{cl}\,W(\underline{\varphi}_{0}^{cl},\underline{n}_{1}^{cl})\,O_{W}(\underline{\varphi}_{M}^{cl},\underline{n}_{M}^{cl})\prod_{m=1}^{M}\delta\left(\underline{\varphi}_{m}^{cl}-\underline{\varphi}_{m-1}^{cl}-\Delta t\,\underline{\nabla}_{n}H_{m-1}(\underline{n}_{m}^{cl},\,\underline{\varphi}_{m-1}^{cl})/\hbar\right)\times\\
\sum_{m^{\prime}=1}^{M-1}\sum_{j=1}^{Ns}\left.\dfrac{\partial^{3}H}{\partial\varphi_{j}^{3}}\right|_{\underline{\varphi}_{m^{\prime}}^{cl}}\delta^{(3)}\left(\underline{n}_{m^{\prime}+1}^{cl}-\underline{n}_{m^{\prime}}^{cl}+\Delta t\,\underline{\nabla}_{\varphi}H(\underline{n}_{m^{\prime}}^{cl},\,\underline{\varphi}_{m^{\prime}}^{cl})/\hbar\right)\prod_{\substack{m=1\\
m\neq m^{\prime}
}
}^{M-1}\delta\left(\underline{n}_{m+1}^{cl}-\underline{n}_{m}^{cl}+\Delta t\,\underline{\nabla}_{\varphi}H_{m}(\underline{n}_{m}^{cl},\,\underline{\varphi}_{m}^{cl})/\hbar\right)\,.
\end{split}
\end{equation}
where the integral over $\varphi_{j,m^{\prime}}^{q}$ was performed
using 
\begin{equation}
\int\ud y\,y^{3}\,e^{-ixy}=i^{3}\dfrac{\partial^{3}}{\partial x^{3}}\int\ud ye^{-ixy}=-i\,2\pi\,\delta^{(3)}(x)\,.
\end{equation}
After a partial integration over $n_{j,m^{\prime}+1}^{cl}$, this
correction term can be expressed as 
\begin{align}
 & \delta\langle\mathcal{O}\rangle_{1}(t)=-i\dfrac{J\rho_{0}a\,\Delta t}{12\,\hbar}\,\sum_{m^{\prime}=1}^{M-1}\sum_{j=1}^{Ns}\,\int\mathcal{D}\varphi^{cl}\,\mathcal{D}n^{cl}\,\sin\varphi_{i,m^{\prime}}^{cl}\,W(\underline{\varphi}_{0}^{cl},\underline{n}_{1}^{cl})\times\nonumber \\
 & \qquad\qquad\qquad\prod_{m=1}^{m^{\prime}}\left[\delta\left(\underline{\varphi}_{m}^{cl}-\underline{\varphi}_{m-1}^{cl}-\Delta t\,\underline{\nabla}_{n}H(\underline{n}_{m}^{cl},\,\underline{\varphi}_{m-1}^{cl})/\hbar\right)\delta\left(\underline{n}_{m+1}^{cl}-\underline{n}_{m}^{cl}+\Delta t\,\underline{\nabla}_{\varphi}H(\underline{n}_{m}^{cl},\,\underline{\varphi}_{m}^{cl})/\hbar\right)\right]\times\nonumber \\
 & \qquad\qquad\qquad\dfrac{\partial^{3}}{\left(\partial n_{j,m^{\prime}+1}^{cl}\right)^{3}}\!\!\left[\prod_{m=m^{\prime}}^{M-1}\!\!\delta\left(\underline{\varphi}_{m+1}^{cl}-\underline{\varphi}_{m}^{cl}-\Delta t\,\underline{\nabla}_{n}H(\underline{n}_{m+1}^{cl},\,\underline{\varphi}_{m}^{cl})/\hbar\right)\right.\times\nonumber \\
 & \qquad\qquad\qquad\left.\prod_{m=m^{\prime}+1}^{M-1}\!\!\!\!\delta\left(\underline{n}_{m+1}^{cl}-\underline{n}_{m}^{cl}+\Delta t\,\underline{\nabla}_{\varphi}H(\underline{n}_{m}^{cl},\,\underline{\varphi}_{m}^{cl})/\hbar\right))\,O_{W}(\underline{\varphi}_{M}^{cl},\underline{n}_{M}^{cl})\right]
\end{align}
which can be rewritten in a more compact form as

\begin{align}
 & \delta\langle\mathcal{O}\rangle_{1}(t)=-i\dfrac{J\rho_{0}a}{12\,\hbar}\,\sum_{j=1}^{Ns}\int_{0}^{t}\ud t^{\prime}\int\ud\underline{\varphi}_{0}^{cl}\,\int\ud\underline{n}_{0}^{cl}\,W(\underline{\varphi}_{0}^{cl},\underline{n}_{0}^{cl})\,\sin\varphi_{j}^{cl}(t^{\prime})\dfrac{\partial^{3}}{\left(\partial n_{j}^{cl}(t^{\prime})\right)^{3}}O_{W}(\underline{\varphi}^{cl}(t),\underline{n}^{cl}(t))\,,\label{eq:correction}
\end{align}
where the trajectories $\underline{\varphi}^{cl}(t^{\prime})$, $\underline{n}^{cl}(t^{\prime})$
are determined by the classical equations of motion \eqref{eq:EOM},
with initial conditions $\{\underline{\varphi}_{0}^{cl},\underline{n}_{0}^{cl}\}$,
just as in the truncated Wigner approximation \eqref{eq:TW}.

The correction term \eqref{eq:correction} can be evaluated by generating
random initial conditions $\underline{\varphi}_{0}^{cl}$ and $\underline{n}_{0}^{cl}$
from the Wigner distribution $W(\underline{\varphi}_{0}^{cl},\underline{n}_{0}^{cl})$
of the initial state, and constructing the classical trajectories
numerically using Eq.~\eqref{eq:EOM}. The functional derivative
with respect to $n_{j}^{cl}(t^{\prime})$ appearing in Eq.~\eqref{eq:correction}
can be determined numerically by adding a small \textquotedbl{}kick\textquotedbl{}
to the trajectory at time $t^{\prime}$: 
\begin{equation}
n_{j}^{cl}(t^{\prime})\rightarrow n_{j}^{cl}(t^{\prime};\varepsilon)=n_{j}^{cl}(t^{\prime})+\varepsilon\,,
\end{equation}
which is then propagated to time $t$ using the equations of motion
\eqref{eq:EOM}. Calculating the modified trajectory $\{\underline{\varphi}^{cl}(t;\varepsilon),\underline{n}^{cl}(t;\varepsilon)\}$
for different kick sizes $\pm\varepsilon$ and $\pm2\,\varepsilon$,
the functional derivative can be evaluated by using the finite difference
expression

\begin{align}
\dfrac{\partial^{3}}{\left(\partial n_{j}^{cl}(t^{\prime})\right)^{3}}O_{W}(\underline{\varphi}^{cl}(t),\underline{n}^{cl}(t))= & \frac{1}{2\,\varepsilon^{3}}\left[O_{W}(\underline{\varphi}^{cl}(t;2\varepsilon),\underline{n}^{cl}(t;2\varepsilon))-O_{W}(\underline{\varphi}^{cl}(t;-2\varepsilon),\underline{n}^{cl}(t;-2\varepsilon))\right.\nonumber \\
 & \quad\quad\left.-2\,O_{W}(\underline{\varphi}^{cl}(t;\varepsilon),\underline{n}^{cl}(t;\varepsilon))+2\,O_{W}(\underline{\varphi}^{cl}(t;-\varepsilon),\underline{n}^{cl}(t;-\varepsilon))\right]\,.\nonumber \\
\end{align}

\begin{figure}[t!]
\centering{}\includegraphics[width=0.85\columnwidth]{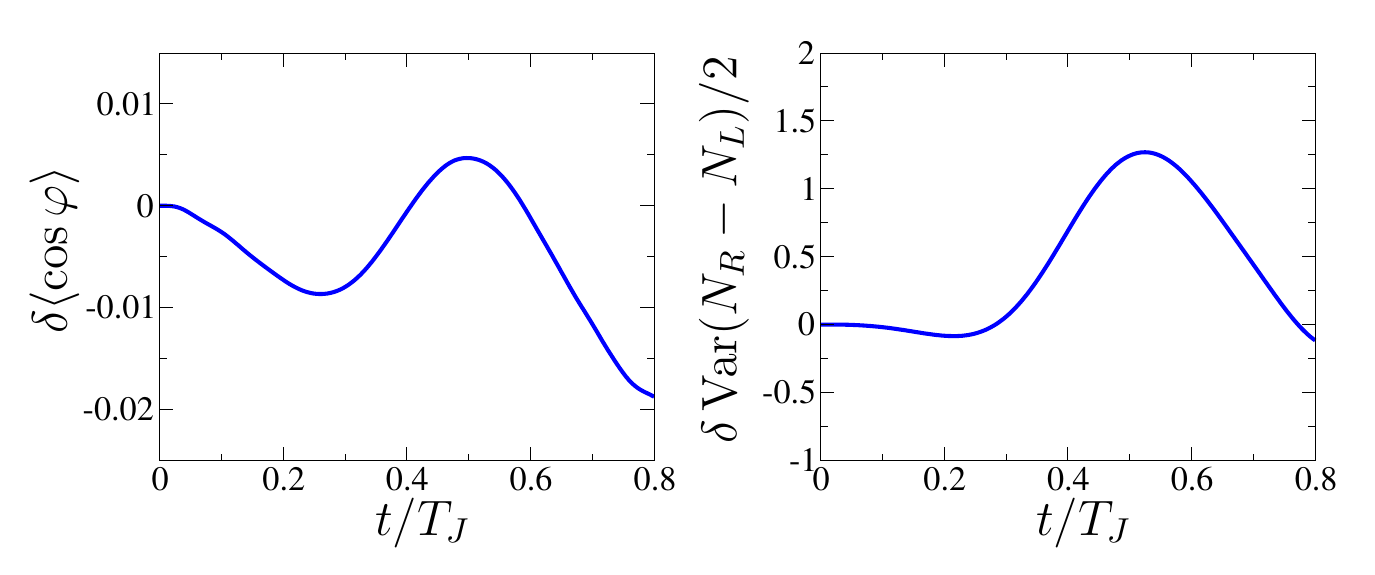}
\caption{First quantum correction to the TWA result. The correction term, Eq.
\eqref{eq:correction}, for the operators $\cos\hat{\varphi}$ and
$(\hat{N}_{R}-\hat{N}_{L})^{2}/4$, plotted as a function of dimensionless
time $f_{J}\,t$, for a quench re-coupling two independent, identical
condensates prepared in their ground state. Here we use the parameters
of Fig. \ref{fig:cos_and_std}: $K=1.56$, $L=14.86\mu m$, $N=400$,
$c=2800\mu m/s$ and $J/h=7Hz$, with the number of lattice sites
$N_{s}=60$. The leading order results for the time evolution of this
quench were analyzed in Sec. \ref{sec:TimeEvolI}. }
\label{fig:twacorr} 
\end{figure}

\begin{figure}[!t]
\centering{}\includegraphics[width=0.4\columnwidth]{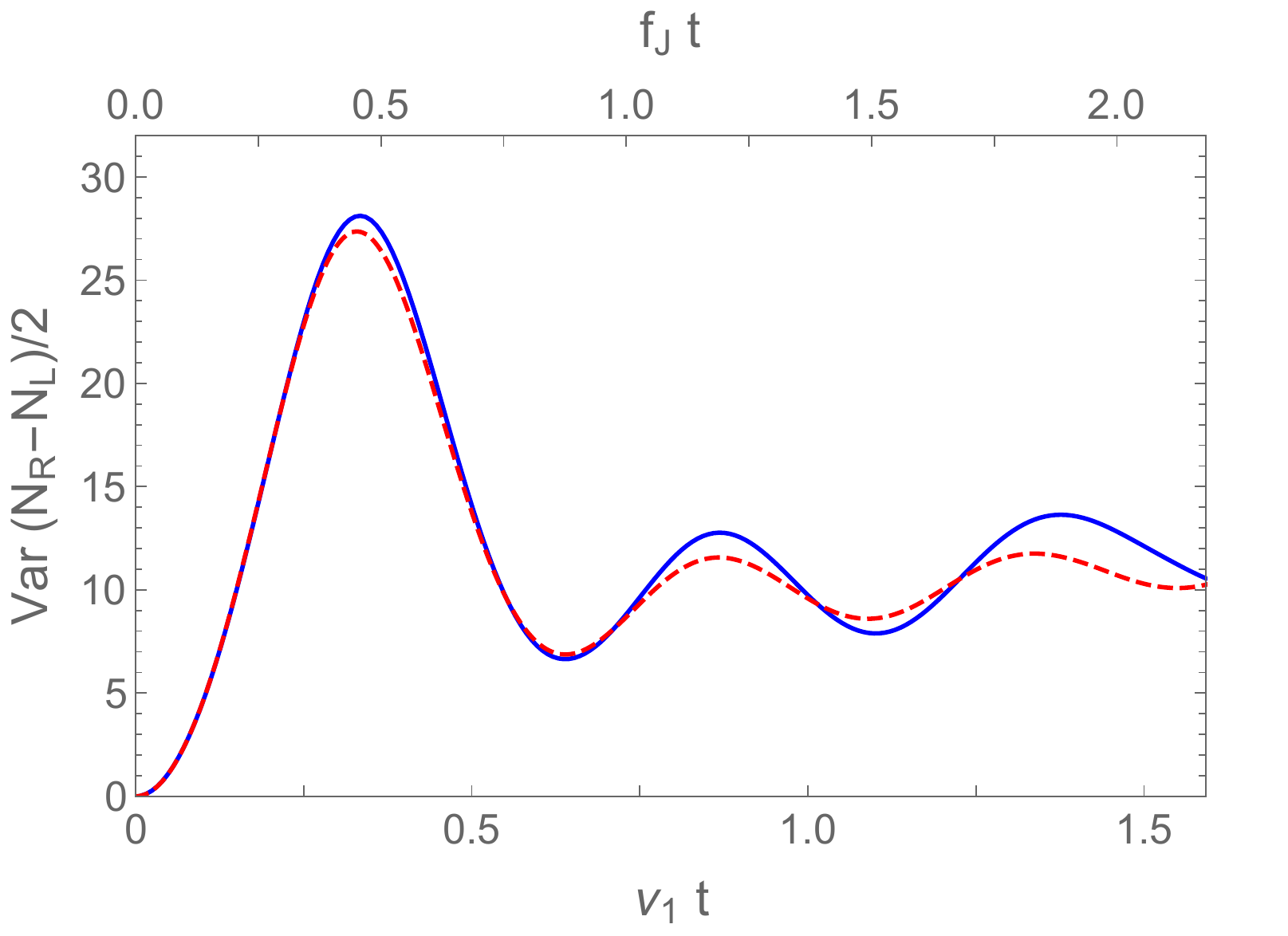}
\caption{Time evolution of the variance of half of the particle number difference,
for the initial state and parameters of Fig. \ref{fig:cos_and_std},
with continuous blue and dashed red curves corresponding to TCSA and
TWA, respectively. The parameters are $K=1.56$, $L=14.86\;\mu{\rm m}$,
$N=400$, $c=2800\;\mu m/s$, $J/h=7{\rm Hz}$ and $N_{s}=60$. Time
is measured in terms of the bare and renormalised Josephson frequencies
$f_{J}$ and $\nu_{1}$ in TWA and TCSA, respectively.}
\label{fig:nvar} 
\end{figure}

The time evolution of the quantum correction term \eqref{eq:correction}
for the operators $\cos\hat{\varphi}$ and $(\hat{N}_{R}-\hat{N}_{L})^{2}/4$
is illustrated in Fig. \ref{fig:twacorr}. Here we considered a quench
already investigated in Sec. \ref{sec:TimeEvolI}, starting with two
independent identical condensates in their ground states and using
the parameters of Fig. \ref{fig:cos_and_std}. In this case the expectation
value of $(\hat{N}_{R}-\hat{N}_{L})^{2}/4$ coincides with the variance
of the particle number difference $(\hat{N}_{R}-\hat{N}_{L})/2$,
because $\langle\hat{N}_{R}-\hat{N}_{L}\rangle=0$ due to left-right
symmetry. Since in Sec. \ref{sec:TimeEvolI} we plotted the standard
deviation of $(N_{R}-N_{L})/2$ instead of the variance, for better
comparison we display Var $(N_{R}-N_{L})/2$ in Fig. \ref{fig:nvar}
for the parameters of Figs. \ref{fig:cos_and_std} and \ref{fig:twacorr}.
By comparing Fig. \ref{fig:twacorr} to Figs. \ref{fig:cos_and_std}
and \ref{fig:nvar}, we find that the quantum correction terms are
not negligible even on quite short time scales compared to the leading
contributions. Nevertheless, the good agreement between the TCSA and
TWA results, demonstrated in Sec. \ref{sec:TimeEvolI}, shows that
the correction term plotted in Fig. \ref{fig:twacorr} considerably
overestimates the error, and the TWA yields a good approximation for
the expectation values of $\cos\hat{\varphi}$ and $(\hat{N}_{R}-\hat{N}_{L})^{2}/4$. 
\end{document}